\documentclass[twocolumn,preprintnumbers,superscriptaddress,nofootinbib,aps,prd,floatfix]{revtex4}

\usepackage{epsf,epsfig,subfigure,graphicx,amsmath,amssymb,cancel}
\usepackage{subfigure}
\usepackage{soul}
\usepackage[colorlinks=true,urlcolor=blue,anchorcolor=blue,citecolor=red,filecolor=blue,linkcolor=blue,menucolor=blue]{hyperref}
\usepackage[usenames,dvipsnames]{xcolor}

\usepackage{afterpage}
\usepackage[utf8]{inputenc}

\allowdisplaybreaks

\newcommand{\newc}{\newcommand}
\newc{\gsim}{\lower.7ex\hbox{$\;\stackrel{\textstyle>}{\sim}\;$}}
\newc{\lsim}{\lower.7ex\hbox{$\;\stackrel{\textstyle<}{\sim}\;$}}
\newc{\gev}{\,{\rm GeV}}
\newc{\mev}{\,{\rm MeV}}
\newc{\ev}{\,{\rm eV}}
\newc{\kev}{\,{\rm keV}}
\newc{\tev}{\,{\rm TeV}}
\newc{\MHT}{$H_T^{\text{miss}}$}
\newc{\MET}{$\slashed{E}_T$}
\newc{\MTT}{$M_{T2}$}

\def\ln{\mathop{\rm ln}}

\newc{\mz}{M_Z}
\newc{\mpl}{M_*}
\newc{\mw}{m_{\rm weak}}
\newc{\nr}[1]{N^c_R{}_{#1}}

\newcommand{\g}{\gamma}

\newcommand{\gbl}{{\rm  g}_{\rm B-L}}

\def\beq{\begin{equation}}
\def\eeq{\end{equation}}
\newcommand{\bea}{\begin{eqnarray}\begin{aligned}}
\newcommand{\eea}{\end{aligned}\end{eqnarray}}
\def\bitem{\begin{itemize}}
\def\eitem{\end{itemize}}

\definecolor{darkgreen}{rgb}{0,0.5,0}
\definecolor{goodyellow}{rgb}{0.9,0.7,0}

\begin{document}
MIT-CTP/5361

\vspace{-1cm}

\title{WIMPs Without Weakness: \\ Generalized Mass Window with Entropy Injection}

\vskip 1.0cm
\author{Pouya Asadi}
\thanks{{\scriptsize Email}: \href{mailto:pasadi@mit.edu}{pasadi@mit.edu}; 
}
\author{Tracy R. Slatyer}
\thanks{{\scriptsize Email}: \href{mailto:tslatyer@mit.edu}{tslatyer@mit.edu};
} 
\affiliation{Center for Theoretical Physics, Massachusetts Institute of Technology, \\ Cambridge, MA, 02139, USA.}
\author{Juri Smirnov}
\thanks{{\scriptsize Email}: \href{mailto:juri.smirnov@fysik.su.se}{juri.smirnov@fysik.su.se}; {\scriptsize ORCID}: \href{http://orcid.org/0000-0002-3082-0929}{ 0000-0002-3082-0929}}
\affiliation{Stockholm University and The Oskar Klein Centre for Cosmoparticle Physics, Alba Nova, \\ 10691 Stockholm, Sweden.}

\begin{abstract}
We study general freeze-out scenarios where an arbitrary number of initial and final dark matter particles participate in the number-changing freeze-out interaction.
We consider a simple sector with two particle species undergoing such a thermal freeze-out; one of the relics is stable and gives rise to the dark matter today, while the other one decays to the Standard Model, injecting significant entropy into the thermal bath that dilutes the dark matter abundance. We show that this setup can lead to a stable relic population that reproduces the observed dark matter abundance without requiring weak scale masses or couplings. The final dark matter abundance is estimated analytically. We carry out this calculation for arbitrary temperature dependence in the freeze-out process and identify the viable dark matter mass and cross section ranges that explain the observed dark matter abundance. This setup can be used to open parameter space for both heavy (above the unitarity bound) or light (sub-GeV) dark matter candidates. We point out that the best strategy for probing most parts of our parameter space is to look for signatures of an early matter-dominant epoch. 

\end{abstract}

\maketitle

\vskip 1cm

\newpage

\section{Introduction}
\label{sec:intro}
The particle nature and non-gravitational interactions of dark matter (DM) are major mysteries of particle physics and cosmology. 
A central classifying feature of a given dark matter scenario is its production mechanism. 
While attractive alternative scenarios, such as freeze-in \cite{Hall:2009bx}, have been put forward recently,
thermal production scenarios via DM freeze-out remain well-motivated.
Perhaps the best-studied production mechanism is the classic thermal freeze-out via 2-to-2 annihilations for Weakly-Interacting Massive Particles (WIMPs)~\cite{Zeldovich:1965gev, Steigman:1984ac, Lee:1977ua}. 
The dark matter yield is given in this case by $Y_{\rm DM} \sim x_{\rm f.o.}/(  m_{\rm DM} M_{\rm Pl} \langle \sigma v \rangle)$, valid as an order of magnitude estimate, as long as the dark sector was in thermal contact with the SM initially \cite{Bringmann:2020mgx}. 
Given the measured temperature of matter-radiation equality $T_{\rm eq}$, the annihilation rate yielding the correct late time dark matter abundance can be determined by setting $T_{\mathrm{eq}} \sim Y_{\mathrm{DM}} m_{\mathrm{DM}} $, implying that:
\bea
\langle \sigma v \rangle \sim  \frac{x_{\rm f.o.}}{T_{\rm eq} M_{\rm Pl}}\,.
\eea

This result has three profound implications. 
First, if $\langle \sigma v \rangle$ remains close to constant after freeze-out, 
there is a fixed target cross section for indirect detection searches. Second, perturbative unitarity provides an upper bound on the DM particle mass. 
Third, given the naive cross section scaling of $\langle \sigma v \rangle \sim \alpha^2/m_{\rm DM}^2$, one expects a relationship between the DM mass scale and the interaction strength of the theory. When this scaling holds, it follows that in this simple scenario:
\begin{equation}
    m_\mathrm{\mathrm{DM}} \sim \alpha \left( T_{\mathrm{eq}} M_\mathrm{Pl}\right)^{1/2} \sim \mathcal{O}(1)~\mathrm{TeV},
    \label{eq:wimpappx}
\end{equation}
where $M_{\mathrm{Pl}}$ is the Planck mass, and $\alpha$ is a coupling constant assumed to be comparable to the coupling of Standard Model (SM) weak interactions. This relation is often referred to as the ``WIMP miracle''; it describes the observation that a weak scale DM mass requires an interaction strength similar to the electroweak interaction. 
\begin{figure}
    \centering
    \resizebox{0.9\columnwidth}{!}{
    \includegraphics{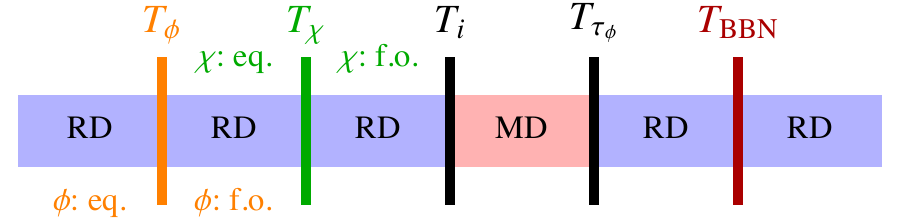}
    }\\
    \resizebox{0.9\columnwidth}{!}{
    \includegraphics{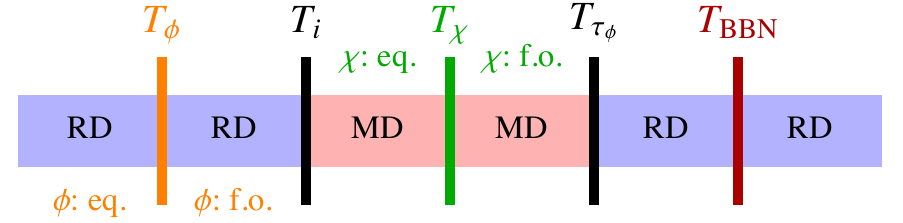}
    }\\
    \caption{Two possible evolutions of the universe in our setup. \textbf{Top (DM freeze-out during a RD epoch):} First, the unstable relic $\phi$ freezes out in a RD epoch at $T=T_\phi$, followed by the DM candidate $\chi$ freezing out at $T=T_\chi$. Then $\phi$ takes over the energy budget of the universe at $T=T_i$ and MD begins.  At $T=T_{\tau_\phi}$, $\phi$ particles decay to SM, diluting the $\chi$ abundance. 
    The decay of $\phi$ should take place before $T_{\mathrm{BBN}}\sim 3$~MeV given the strong bounds \cite{Jedamzik:2006xz,Kawasaki:2017bqm,Sabti:2019mhn,Depta:2019lbe}. \textbf{Bottom (DM freeze-out during a MD epoch):} Similar timeline as above, but now $T_\chi \leqslant T_i$, i.e. the universe enters the early MD epoch before $\chi$ freezes out, which affects its final abundance. 
    }
    \label{fig:evolution}
\end{figure}

The interaction topology of the freeze-out process can be generalized to the case where DM depletion is dominated by number-changing interactions with $p$ DM particles in the initial state and $q$ DM particles in the final state, with $p>q$; see e.g. Ref.~\cite{Hochberg:2014dra} for a $3 \rightarrow 2$ freeze-out scenario known as Strongly Interacting Massive Particle (SIMP) freeze-out.  Parametrizing the interaction rate factor $\langle \sigma v^{p-1} \rangle \sim \alpha^p/m_{\rm DM}^{3p-4}$, requiring the correct relic density relates the mass and interaction strength as 
\begin{equation}
    m_\mathrm{\chi} \sim 
    \alpha \, T_{\mathrm{eq}} \left(\frac{M_\mathrm{Pl}}{T_{\mathrm{eq}}}\right)^{1/p}  \sim \alpha \, 10^{28/p-9}~\mathrm{GeV}\, .
    \label{eq:simpappx}
\end{equation}
For $p > 2$ the DM mass is constrained to be in the sub-GeV range; for such a light DM relic, 
strong astrophysical/cosmological bounds exist due to DM self interaction constraints \cite{Bondarenko:2020mpf} and the verified predictions of Big Bang Nucleosynthesis (BBN) \cite{Jedamzik:2006xz,Kawasaki:2017bqm,Sabti:2019mhn}. 

Furthermore, if multiple dark-sector species are produced by similar freeze-out mechanisms, but one or more species decays with a timescale shorter than the age of the universe, then the decays of the unstable components can inject significant entropy into the visible sector and modify the yield of any remaining dark components. We consider a simple two-state version of this scenario, where an already frozen out relic $\chi$ is diluted by entropy injection from another relic $\phi$ (following the notation of Ref.~\cite{Bramante:2017obj}). Possible timelines for this scenario are shown in Fig.~\ref{fig:evolution}. To obtain a non-negligible entropy injection, $\phi$ should be both abundant and long-lived enough to dominate the energy budget of the universe, giving rise to an early matter-dominant (MD) epoch starting at temperature $T_i$ \cite{Cirelli:2018iax}.

While $\phi$ always freezes out during a radiation-dominant (RD) epoch, for some points of the parameter space $\chi$'s freeze-out can happen after $\phi$ dominates the energy density of the universe, i.e. during a MD epoch (bottom figure in Fig.~\ref{fig:evolution}); see Ref.~\cite{Hamdan:2017psw} for a previous work on DM freeze-out during an early MD epoch. 

We study the freeze-out of both $\chi$ and $\phi$ from the thermal bath when they are controlled by general interactions beyond the conventional WIMP and SIMP scenarios. These freeze-outs happen, respectively, at temperatures $T_\chi$ and $T_\phi$. 
We analytically calculate the final abundance of $\chi$ after the entropy injection and will show that such a minimal extension can vastly expand the viable parameter space. We also check our analytic formulas against numerical results for a few benchmark scenarios and find better than $10 \%$ agreement in the final relic abundance calculation.

Given the vastly different search strategies required to look for DM in different parts of the parameter space, identifying new dynamics that can expand the viable parameter space can have profound implications for the experimental efforts. 
We find that our minimal extension can substantially affect the viable mass range of thermal relics. We will argue that the best way to probe the high mass end of our parameter space is to look for signatures of an early MD epoch, as discussed in Refs.~\cite{Erickcek:2011us,StenDelos:2019xdk,Barenboim:2021swl}.

The effects of entropy injection in the early universe have been considered previously in the literature. It has been shown that in a setup with DM particles freezing out during a RD epoch and with 2-to-2 interactions controlling the freeze-out, the DM mass can be as large as $10^{10}$~GeV \cite{Bramante:2017obj}. If the baryon asymmetry of the universe (BAU) can be generated after the dilution, even higher masses are permitted.
The effect of such a dilution on non-thermal \cite{Allahverdi:2018aux} or hot thermal relics \cite{Evans:2019jcs} has also been studied in the literature. 
Entropy injection has been studied in setups where the DM and SM have different temperatures and where the 2-to-2 interactions controlling DM freeze-out are $\chi \chi \rightarrow \phi \phi$~\cite{Berlin:2016gtr}. The decay of inflatons to decoupled SM and DM particles (without freeze-out or freeze-in) has been studied in Ref.~\cite{Randall:2015xza}.

In this work, we extend these studies by analytically studying general number-changing interactions for both $\phi$ and $\chi$ freeze-outs, including the case with $p \geqslant 4$ initial particles. We also consider general temperature dependence in the freeze-out cross sections and the effect of $\chi$ freezing out during an early MD epoch. 
We will use the result of our analytic calculations to demonstrate the following:
\begin{itemize}
    \item The setup naturally opens up parameter space between the freeze-in and freeze-out DM annihilation cross sections.
    \item Lower interaction rates arise due to small couplings between DM and SM particles for DM masses at or below the GeV scale. We thus find new open parameter space for sub-GeV DM masses, which can be tested in the future by direct detection experiments with sensitivity to very-low-energy recoils, and improved indirect detection searches.
    \item Even with $\phi$ lifetimes as short as the electroweak phase transition timescale, it is possible to have a DM candidate with a mass that exceeds the WIMP unitary bound~\cite{Griest:1989wd} of $\sim 100 \, \rm TeV$, and with unexpectedly large coupling values $\alpha \sim \mathcal{O}(1)$.
    \item In scenarios with $3\rightarrow 2$ freeze-out interactions we find new parameter space with masses far above the GeV scale. 
    \item Finally, we show that in general DM freeze-out via interactions with $p \geqslant 4$ are viable even for heavy DM relics.
\end{itemize}

In Sec.~\ref{sec:dilution} we summarize the result of our analytic calculation for entropy injection and general freeze-out mechanisms. In Sec.~\ref{sec:parameters} we use our analytic calculation to identify the viable range of cross sections and masses that give rise to the observed DM abundance today for a few different benchmark freeze-out interactions. 
In Sec.~\ref{sec:experiments} we show the parameter space that entropy injection opens in a sample minimal model with a $Z'$ mediator. 
We conclude in Sec.~\ref{sec:conclusion}. 
Further details on the entropy injection calculation, freeze-out during a RD epoch, and freeze-out during a MD epoch are provided in Apps.~\ref{app:entropydump}-\ref{app:MD}, respectively. We also comment more on a few other benchmark freeze-out scenarios in App.~\ref{app:morescenarios}.

\section{freeze-out and Dilution in a Minimal Entropy injection Setup}
\label{sec:dilution}

\subsection{Entropy injection}
\label{subsec:entropy}
Consider a minimal DM setup with a single DM candidate $\chi$, with mass $m_{\chi}$, that freezes out. Unless specified otherwise, we will take $\chi$ to be self-conjugate (e.g. a Majorana fermion or real scalar). Throughout this work, we assume that kinetic equilibrium is maintained between the dark sector particles and the SM. 
If the interaction controlling the freeze-out has $p_\chi=2$ DM particles in the initial, $q_\chi=0$ DM particles, and two SM particles in the final state, the relic abundance will be given by \cite{Kolb:1990vq}
\begin{equation}
    \Omega h^2 \approx 3.81 \times \frac{s_0}{\rho_c}  h^2   \frac{x_{\mathrm{f.o.},\chi} g_{*,\chi}^{1/2}}{g_{*S,\chi}} \frac{1}{\sigma_{0,p_\chi} M_{\mathrm{Pl} }}   .
    \label{eq:22KolbTurner}
\end{equation}
Here $s_0$ and $\rho_c$ are today's entropy density and critical energy density, respectively, and the ratio has a numerical value of \cite{Zyla:2020zbs}
\begin{equation}
    \frac{s_0}{\rho_c}h^2 \approx 2.75 \times 10^8 ~\mathrm{GeV}^{-1},
    \label{eq:s0rhoc}
\end{equation}
$x_{\mathrm{f.o.},\chi}=m_\chi/T_\chi$ with $T_\chi$ being the freeze-out temperature, $g_{*(S),\chi}$ is number of relativistic degrees of freedom (for entropy) at $T=T_\chi$, $M_\mathrm{Pl}\approx 1.2 \times 10^{19}$~GeV is the Planck mass, and we used $\langle \sigma v \rangle \equiv \sigma_{0,p_\chi}$ as the cross section for the interaction controlling the freeze-out, i.e. freeze-out interactions do not have any temperature dependence. 
Using this, it has been argued that the natural mass range for a DM particle freezing out via $p_\chi=2$ interactions is around a TeV, with an upper bound of $\sim 10^2$ TeV on its mass owing to unitarity arguments \cite{Griest:1989wd}.

However, there are a number of ways to break outside this mass window. One simple possibility is when, in addition to DM and its portal to the SM, there is another relic $\phi$ with mass $m_\phi$ that can freeze out and at some point decay to SM particles, injecting a significant amount of entropy into the SM thermal bath. For this to happen, the unstable relic $\phi$ should take over the energy budget of the universe after its freeze-out \cite{Kolb:1990vq}, giving rise to an early MD epoch. If such an epoch starts at $T=T_i$, it can be shown that the injected entropy will dilute the DM relic abundance by a factor:
\begin{equation}
    \xi \equiv \left(\frac{S_f}{S_i}\right)^{-1}  = \left( 1+ 1.65 <g_{*S}^{1/3}> \left(\frac{T_i^4}{(\Gamma_\phi M_{pl} )^{2}} \right)^{1/3} \right)^{-3/4}.
    \label{eq:EntDump}
\end{equation}
where $S_f$ ($S_i$) is the SM bath total entropy after (before) the entropy injection, $\Gamma_\phi$ is $\phi$'s decay rate, and $\langle g_{*}^{1/3} \rangle$ is a weighted average over relativistic degrees of freedom throughout the MD epoch; see App.~\ref{app:entropydump} for further details. We assume $\phi$ can only decay to SM bath particles. The effect of this entropy injection on the DM relic abundance with $p_\chi=2,~q_{\chi}=0$ (conventional WIMP) has been studied in Ref.~\cite{Bramante:2017obj}.

We extend the work of Ref.~\cite{Bramante:2017obj} to the general case of arbitrary $p_\chi$ and $q_\chi$, and the assumption that the abundance of $\phi$ itself is set by a freeze-out process.  Furthermore, while the dilution factor $\xi$ in Ref.~\cite{Bramante:2017obj} was treated as a free parameter itself, we consider scenarios where $\phi$ freezes out via a process with $p_\phi$ and $q_\phi$ particles in the initial and final states respectively. We also consider a general temperature dependence for the freeze-out cross section
\begin{equation}
    \langle \sigma v^{p_\phi -1} \rangle \equiv \sigma_{0,p_\phi} x_\phi^{-l_\phi}.
    \label{eq:xsecphi}
\end{equation}
In the conventional case of $p_\phi=2$ interactions, $l_\phi$ corresponds to the partial wave expansion of the interaction in the non-relativistic regime ($l_\phi=0$ for $s$-wave, $l_\phi=1$ for $p$-wave, etc.); for higher $p_\phi$ interactions the situation is somewhat more complicated. For simplicity, and to further capture the effect of other phenomena such as Sommerfeld enhancement \cite{Hisano:2004ds,Hisano:2006nn,Cirelli:2007xd,Arkani-Hamed:2008hhe,Cassel:2009wt,Slatyer:2009vg}, we remain agnostic about the interpretation of $l_\phi$ and merely use it to capture the temperature-dependence of the freeze-out interaction. 
In App.~\ref{app:RD} we show that
\begin{eqnarray}
\label{eq:TiRDMD}
T_i &=& \frac{30 }{\pi^2} \frac{g_{*i}}{g_{*Si}} \left(\frac{p_\phi !}{(p_\phi-q_\phi)(p_\phi-1)}\right)^{\frac{1}{p_\phi-1}}  \left(   1.67 g_{*,\phi}^{1/2}  \right)^{\frac{1}{p_\phi-1}} \nonumber \\ 
& \times &\frac{m_\phi^{\frac{2}{p_\phi-1}-2}}{g_{*S,\phi}} \left( \frac{(3p_\phi+l_\phi-5)x_{\mathrm{f.o.},\phi}^{3p_\phi+l_\phi-5} }{\sigma_{0,p_\phi} M_{\mathrm{Pl}} } \right)^{\frac{1}{p_\phi-1}}, 
\end{eqnarray}
where $x_{\mathrm{f.o.},\phi}=m_\phi/T_\phi$ with $T_\phi$ being $\phi$'s freeze-out temperature, $g_{*(S)i}$ is the relativistic degrees of freedom (for entropy) at $T=T_i$, and $g_{*(S),\phi}$ is the relativistic degrees of freedom (for entropy) at $\phi$'s freeze-out. In our setup, where all the particles are at the same temperature, we can simplify the equation further with $g_{*}=g_{*S}$ at all temperatures. In deriving this equation we also assumed $\phi$ is identical to its anti-particle.

Next, we should study the freeze-out of $\chi$, distinguishing between the RD and MD epochs in the two following sections.

\subsection{Freeze-out during a RD epoch}
\label{subsec:FOinRD}

We consider interactions where $p_\chi$ DM particles in the initial state convert to $q_\chi$ DM particles ($p_\chi \geqslant q_\chi+1$) and an arbitrary number of SM particles in the final state. 
We also assume for both $\chi$ and $\phi$ particles and anti-particles are identical; this affects the symmetry factors used in the equations. 

The Boltzmann equation for such relics will be
\begin{equation}
s\dot{Y} = - \frac{(p-q)}{p!} \langle \sigma v^{p-1}\rangle s^p Y^p \left(1 - \left(\frac{Y_{\mathrm{eq}}}{Y}\right)^{p-q}\right),
\label{eq:BoltzEqpq}
\end{equation}
where $p=p_\chi,~p_\phi$ ($q=q_\chi,~q_\phi$) refers to the initial (final) number of $\chi$ or $\phi$ particles involved in their respective freeze-out processes, $s$ is the SM entropy density, $Y$ is the relic's yield, and $\langle \sigma v^{p-1} \rangle$ is the relevant cross section. 
Notice that the number of SM particles involved in the interaction does not enter the equation. Solving this equation for relic $\chi$, we find (see App.~\ref{app:RD} for further details):
\begin{eqnarray}
\label{eq:finalOmegaRD}
\Omega_\chi h^2 &\approx&  \frac{45 }{2\pi^2} \frac{s_0}{\rho_c} h^2 \left(\frac{p_\chi !}{(p_\chi-q_\chi)(p_\chi-1)}\right)^{\frac{1}{p_\chi-1}}  \\ 
&\times & \left(   1.67 g_{*,\chi}^{1/2}  \right)^{\frac{1}{p_\chi-1}} 
\times \frac{m_\chi^{\frac{2}{p_\chi-1}-2}}{g_{*S,\chi}} \nonumber \\
&  \times & \left( \frac{(3p_\chi+l_\chi-5)x_{\mathrm{f.o.},\chi}^{3p_\chi+l_\chi-5} }{\sigma_{0,p_\chi} M_{\mathrm{Pl}} } \right)^{\frac{1}{p_\chi-1}}, \nonumber
\end{eqnarray}
where the interaction cross section is defined as
\begin{equation}
    \langle \sigma v^{p_\chi-1} \rangle \equiv \sigma_{0,p_\chi} x_\chi^{-l_\chi},
    \label{eq:xsecchi}
\end{equation}
with $\sigma_{0,p_\chi}$ a prefactor, $x_{\chi} = m_\chi/T$, and $l_\chi$ capturing the temperature dependence of the interaction. 
(A slightly more general result can be found in App.~\ref{app:RD}, see Eq.~\eqref{eq:solonepartial}.) 
We can easily check that in the limit of $p_\chi =2,~q_\chi=0$, Eq.~\eqref{eq:finalOmegaRD} reduces to Eq.~\eqref{eq:22KolbTurner}. 
As a cross check, one can verify that this expression has the right dimensions; in doing so, it should be noted that $\sigma_{0,p_\chi}$ has mass dimension $-3p_\chi+4$. A similar calculation can be repeated for the freeze-out of $\phi$ and its abundance before its decay. We checked Eq.~\eqref{eq:finalOmegaRD} explicitly against numerical calculations for a few benchmark scenarios and found better than $10\%$ agreement.

Combining Eq.~\eqref{eq:finalOmegaRD} with Eqs.~\eqref{eq:EntDump}-\eqref{eq:TiRDMD}, we find an expression for the $\chi$ relic abundance after the entropy injection 
\begin{widetext}
\begin{eqnarray}
\Omega_\chi  h^2 & \approx &  0.45 \times \frac{s_0}{\rho_c} h^2 \times \left(	1.67	\right)^\frac{p_\phi-p_\chi}{(p_\chi-1)(p_\phi-1)}  \times \frac{\left(	p_\chi !	\right)^\frac{1}{(p_\chi-1)}}{\left(	p_\phi !	\right)^\frac{1}{(p_\phi-1)}}  
 \times  \frac{      \left(   (p_\phi - q_\phi ) (p_\phi - 1 )  \right)^{\frac{1}{p_\phi-1}}     }{\left(   (p_\chi - q_\chi ) (p_\chi - 1 )  \right)^{\frac{1}{p_\chi-1}}  }  \times \frac{  \left(  l_\chi+3p_\chi-5 \right)^{\frac{1}{p_\chi-1}} }{   \left( l_\phi+3p_\phi-5  \right)^{\frac{1}{p_\phi-1}}   } \nonumber \\ 
\label{eq:finalpqdilutedresult}
&\times & \frac{  x_{\mathrm{f.o.},\chi}^{\frac{l_\chi+3p_\chi-5}{p_\chi-1}}   }{  x_{\mathrm{f.o.},\phi}^{\frac{l_\phi+3p_\phi-5}{p_\phi-1}}  } 
 \times   \langle g_{*S,\phi}^{1/3} \rangle^{-3/4}  \times \frac{g_{*S,\phi}}{g_{*S,\chi}} \times \frac{g_{*,\chi}^{\frac{1}{2(p_\chi-1)}}}{g_{*,\phi}^{\frac{1}{2(p_\phi-1)}}}  
 \times    M_{\mathrm{Pl}}^{\frac{1}{2}+ \frac{p_\chi-p_\phi}{(p_\chi-1)(p_\phi-1)} } \times \frac{m_\phi^{2 \frac{p_\phi-2}{p_\phi-1}}  }{m_\chi^{2 \frac{p_\chi-2}{p_\chi-1}}}    \times       \frac{\sigma_{0,p_\phi}^{\frac{1}{p_\phi-1}}  }{\sigma_{0,p_\chi}^{ \frac{1}{p_\chi-1}}}                 \times \Gamma_\phi^{\frac{1}{2}}.
\end{eqnarray}
\end{widetext}
The following assumptions were used to arrive at this equation:
\begin{enumerate}
    \item both $\phi$ and $\chi$ freeze out during a RD epoch,
    \item the entropy injection happens after $\chi$ has frozen out from the thermal bath, and
    \item we neglected the factor of 1 in Eq.~\eqref{eq:EntDump} for $\xi$, i.e. we assumed $\xi \ll 1$.
\end{enumerate}
Furthermore, as argued in Ref.~\cite{Bramante:2017obj}, if the BAU is generated before $\phi$ decays, we should check that $\xi \gsim \eta_\mathrm{b}$ with $\eta_\mathrm{b}=6 \times 10^{-10}$ \cite{Zyla:2020zbs} denoting the observed baryon-to-photon ratio; otherwise any pre-existing asymmetry will be washed out to values too small to explain today's BAU. However, if a baryogenesis mechanism is provisioned for after the entropy injection, there is no obstruction to going to smaller values of $\xi$. 

Equation~\eqref{eq:finalpqdilutedresult} shows that $\Omega_\chi$ is most sensitive to $p_{\phi,\chi}$ (since they appear in powers of various quantities), masses, freeze-out cross sections, and $\phi$'s decay rate to SM $\Gamma_\phi$; other parameters mostly affect the relic abundance by $\mathcal{O}(1)$ factors. 
Unlike the case of $p_\chi=2$ we find that the final abundance does explicitly depend on $m_\chi$ for $p_\chi \geqslant 3$. 
We also find that the final abundance always increases as $\sigma_{0,p_\chi}$ decreases. 
We can also calculate $x_{\mathrm{f.o.},\chi}$ as a function of other parameters, see App.~\ref{app:RD}. 
Notice that $\Gamma_\phi$ is a free parameter of the calculation at this level. By varying this parameter, it is possible to go to different regions of the parameter space with different DM masses. In particular, the entropy injection dilutes the DM abundance, which can be used to open up new parameter space for DM masses beyond the unitarity bound \cite{Griest:1989wd}. We will discuss this in more detail in the upcoming section. 

\subsection{Freeze-out during a MD epoch}
\label{subsec:FOinMD}

When there is a large hierarchy between the masses of the two relics, $\phi$ takes over the energy budget of the universe and the universe enters an early MD epoch before $\chi$ freezes out, i.e. $T_\chi \leqslant T_i$, see the bottom row of Fig.~\ref{fig:evolution}.

Here $\phi$'s freeze-out goes forward as before. We can check from the calculation of App.~\ref{app:RD} that its asymptotic yield $Y_\infty^\phi$ will be given by
\begin{eqnarray}
\label{eq:solallpartial3repeatMD}
Y_\infty^\phi &=& \frac{45 }{2\pi^2} \left(\frac{p_\phi !}{(p_\phi-q_\phi)(p_\phi-1)}\right)^{\frac{1}{p_\phi-1}}  \left(   1.67 g_{*\phi}^{1/2}  \right)^{\frac{1}{p_\phi-1}} \nonumber \\
&\times & \frac{m_\phi^{\frac{2}{p_\phi-1}-3}}{g_{*S,\phi}} \left( \frac{(3p_\phi+l_\phi-5)x_{\mathrm{f.o.},\phi}^{3p_\phi+l_\phi-5}}{\sigma_{0,p_\phi} M_{\mathrm{Pl}}  } \right)^{\frac{1}{p_\phi-1}} ,
\end{eqnarray}
and its energy density can be written as 
\begin{eqnarray}
\label{eq:MDrhophi}
\rho_\phi (T) &=& m_\phi Y_\infty^\phi s(T) \approx \frac{\pi^2}{30} g_{*S,\phi} T_i T^3,
\end{eqnarray}
where $s(T)$ is the SM entropy at temperature $T$. In the second equality we neglect the change in $g_{*S}$ after the $\phi$ freeze-out and use Eq.~\eqref{eq:TiRDMD}. This was done to simplify the temperature-dependence of $\rho_\phi (T)$ so as to simplify the upcoming integrations. We will see eventually that as long as we do not introduce any new degrees of freedom this will only introduce $\mathcal{O}(1)$ change in the final DM abundance expression. 

In App.~\ref{app:MD} we provide more details on the Boltzmann equations governing $\chi$ freeze-out during this MD epoch. 
The final result for the $\chi$ abundance before the entropy injection is
\begin{eqnarray}
\label{eq:finalOmegaMD}
\Omega_\chi h^2 &=& \frac{45 }{2\pi^2} \frac{s_0}{\rho_c}h^2 \left(\frac{p_\chi !}{(p_\chi-q_\chi)(p_\chi-1)}\right)^{\frac{1}{p_\chi-1}}   \\
&\times & \left(    
 1.67 g_{*S,\phi}^{1/2} T_i^{1/2} \right)^{\frac{1}{p_\chi-1}} 
 \times  \frac{m_\chi^{\frac{3/2}{p_\chi-1}-2}}{g_{*S_\chi}} \nonumber \\
  &\times & \left( \frac{(3p_\chi+l_\chi-9/2)x_{\mathrm{f.o.},\chi}^{3p_\chi+l_\chi-9/2} }{\sigma_{0,p_\chi} M_{\mathrm{Pl}} } \right)^{\frac{1}{p_\chi-1}}. \nonumber
\end{eqnarray}
Comparing this with Eq.~\eqref{eq:finalOmegaRD} we observe changes in the power of $x_{\mathrm{f.o.}}$ and DM mass, as well as the appearance of $T_i$ in the calculation. 

We checked this equation against the numerical result as well and found very good agreement. Using Eq.~\eqref{eq:EntDump} and Eq.~\eqref{eq:TiRDMD} for the entropy injection, we find an expression for the final DM abundance today: 
\begin{widetext}
\begin{eqnarray}
\label{eq:finalpqdilutedresultMD}
 \Omega_\chi  h^2  &\approx &  0.45 \times \frac{s_0}{\rho_c}h^2  \times           \left(	1.67	\right)^\frac{p_\phi-p_\chi}{(p_\chi-1)(p_\phi-1)}                    \times \frac{\left(	p_\chi !	\right)^\frac{1}{(p_\chi-1)}}{\left(	p_\phi !	\right)^\frac{1}{(p_\phi-1)}}
 \times  \frac{      \left(   (p_\phi - q_\phi ) (p_\phi - 1 )  \right)^{\frac{1}{p_\phi-1}}     }{\left(   (p_\chi - q_\chi ) (p_\chi - 1 )  \right)^{\frac{1}{p_\chi-1}}  }  \times \frac{  \left(  l_\chi+3p_\chi-\frac{9}{2} \right)^{\frac{1}{p_\chi-1}} }{   \left( l_\phi+3p_\phi-5  \right)^{\frac{1}{p_\phi-1}}   }  \\
&\times & \frac{  x_{\mathrm{f.o.},\chi}^{\frac{l_\chi+3p_\chi-\frac{9}{2}}{p_\chi-1}}   }{  x_{\mathrm{f.o.},\phi}^{\frac{l_\phi+3p_\phi-5}{p_\phi-1}}  } \times    \langle g_{*S,\phi}^{1/3} \rangle^{-3/4}   \times \frac{g_{*S,\phi}}{g_{*S,\chi}} \times \frac{g_{*S,\phi}^{\frac{1}{2(p_\chi-1)}}}{g_{*,\phi}^{\frac{1}{2(p_\phi-1)}}} \times T_i^{\frac{1}{2(p_\chi-1)}}
 \times     M_{\mathrm{Pl}}^{\frac{1}{2}+ \frac{p_\chi-p_\phi}{(p_\chi-1)(p_\phi-1)} } \times \frac{m_\phi^{2 \frac{p_\phi-2}{p_\phi-1}}  }{m_\chi^{2 \frac{p_\chi-7/4}{p_\chi-1}}}    \times       \frac{\sigma_{0,p_\phi}^{\frac{1}{p_\phi-1}}  }{\sigma_{0,p_\chi}^{ \frac{1}{p_\chi-1}}}                 \times \Gamma_\phi^{\frac{1}{2}}, \nonumber
\end{eqnarray}
\end{widetext}
We should reiterate that in deriving this equation we neglected the evolution of $g_{*S}$ between $T_\phi$ and $T_\chi$ for analytic clarity; this will merely introduce $\mathcal{O}(1)$ effects on the DM abundance calculation. This dependence should be restored in a fully numerical treatment. It should also be noted that $T_i$ introduces some further dependence on $m_\phi$ and $\sigma_{0,\phi}$, see Eq.~\eqref{eq:TiRDMD}.

\section{Viable DM Parameter Space}
\label{sec:parameters}

We can use Eqs.~\eqref{eq:finalpqdilutedresult} and \eqref{eq:finalpqdilutedresultMD} to study the available parameter space in models with different freeze-out dynamics. In this section, we consider a few benchmark scenarios and comment on the new parameter space opened thanks to the entropy injection.

\subsection{Broadening the thermal cross section range}
\label{subsec:www}

Let us start by considering the WIMP-like freeze-out case of $p_{\chi,\phi}=2,~q_{\chi , \phi}=0$. In this setup, if the freeze-outs happen during a RD epoch the final relic abundance does not explicitly depend on the masses.
Without any entropy injection the observed DM abundance would require 
\begin{equation}
    \langle \sigma v \rangle_{\mathrm{WIMP}} \sim 2-3 \times 10^{-26}\,\mathrm{cm}^3/\mathrm{s}.
    \label{eq:xsecWIMP}
\end{equation}
In our scenario, however,  no absolute cross section or mass scale is singled out. The final abundance is instead determined by the ratio of the $\chi$ and $\phi$ annihilation cross sections. 
After the entropy injection, the right relic abundance is obtained even with cross section values below the standard WIMP thermal cross section. 

\begin{figure}
    \centering
    \resizebox{0.9\columnwidth}{!}{
    \includegraphics{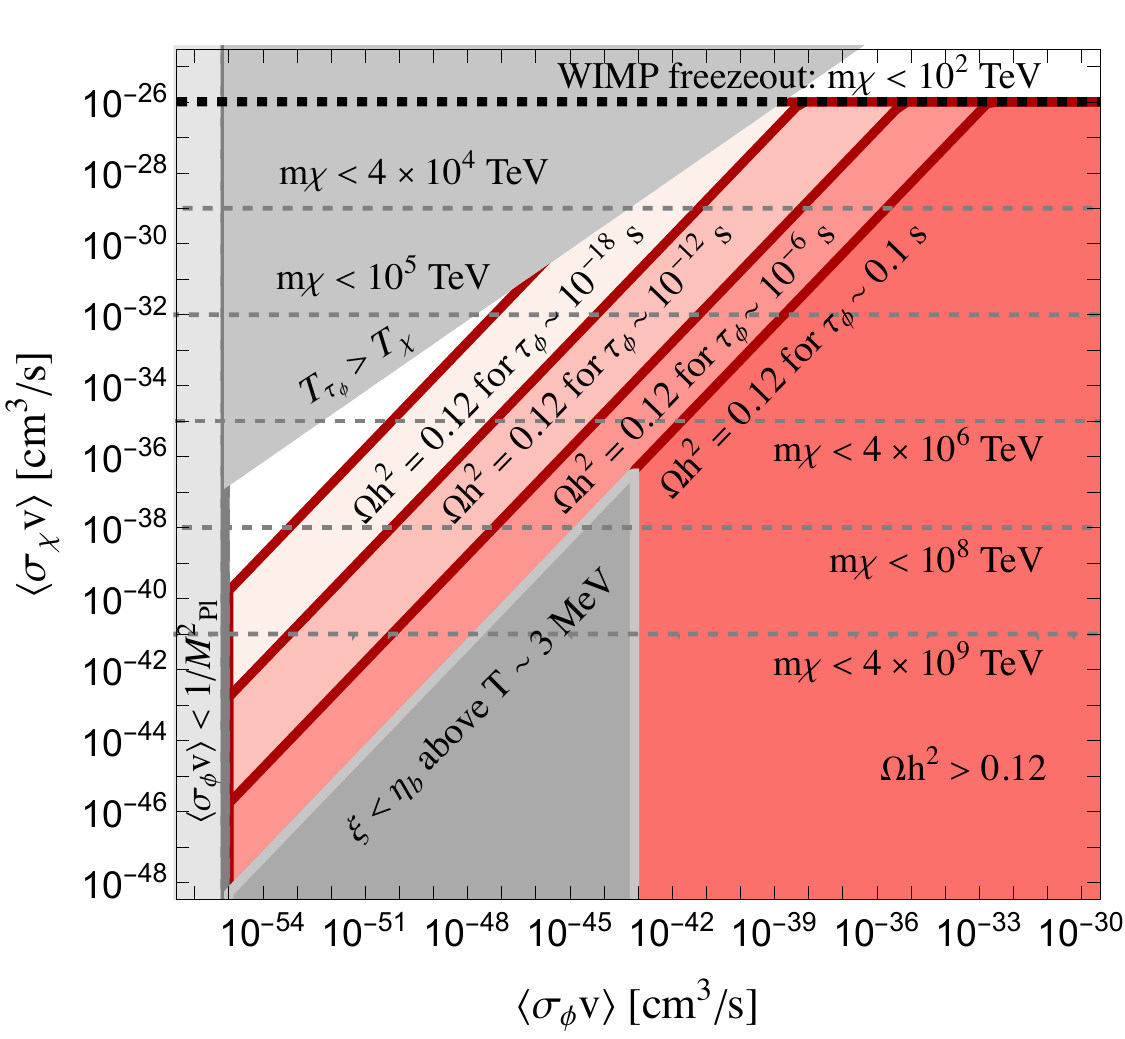} 
    }\\
    \caption{The viable range of cross sections for a $2 \rightarrow 0$ freeze-out process for both $\phi$ and $\chi$. Using the unitarity bound the cross sections can be associated with an upper bound on the mass of the particle, denoted by dashed gray lines. On the red lines (below the red lines) we can get the observed relic abundance (overclose the universe) by assuming the denoted lifetime for $\phi$. In the upper left corner of the parameter space the dilution factors will always be too small (for $T_{\tau_\phi} \leqslant T_\chi$) and we underclose the universe. For the points in the lower gray region, the dilution factor will be small enough that the BAU is completely washed away at or after BBN. For $\tau_\phi \sim 0.1$~s there is no time left to reproduce this asymmetry before BBN; yet, for smaller lifetimes the BAU can be generated after $\phi$'s decay.  }
    \label{fig:WWparameters}
\end{figure}

In Fig.~\ref{fig:WWparameters} we show the  annihilation cross section values for $\chi$ and $\phi$ that yield the correct DM abundance, at different lifetimes for $\phi$. We use $l_{\chi,\phi}=0$ ($s$-wave interactions) in generating this plot, and assume both relics freeze out during a RD epoch. 
We assume a geometric $s$-wave cross section, 
\begin{equation}
    \langle \sigma_\chi v \rangle \lesssim  \frac{4\pi}{m_\chi^2}.
    \label{eq:unitaritysigmap2}
\end{equation}
This allows us to relate each $\langle \sigma_\chi v \rangle$ to a maximal DM mass. This should be viewed  as an approximate upper limit for the $s$-wave cross section during freeze-out, for a given $m_\chi$, since the unitarity bound on the non-relativistic $s$-wave cross section is $\langle \sigma v \rangle < 4\pi/k^2$ and freeze-out occurs when the DM is only mildly non-relativistic (i.e. $k$ is parametetrically similar to $m_\chi$). The $m_\chi$ values obtained by this relation can thus also be viewed as approximate upper bounds on the DM mass for a fixed cross section.

An additional factor to consider is the baryon to photon ratio of the universe. If the dilution factor $\xi$ becomes smaller than around $\eta_{\mathrm{b}} \approx 6 \times 10^{-10}$, any previously-produced baryon asymmetry dilutes to values smaller than today's observed asymmetry, unless $\phi$ decays generate a SM baryon asymmetry.  Thus, for this part of the parameter space, a baryogenesis mechanism should be provisioned that is active after the $\phi$ entropy injection. 

In general we will consider lifetimes $\tau_\phi \lesssim 0.1$~s; this choice ensures the bulk of the energy density stored in $\phi$ has decayed by BBN. A more careful treatment of BBN constraints may open a small parameter space for larger lifetimes, see Refs.~\cite{Jedamzik:2006xz,Kawasaki:2017bqm}; here we use $\tau_\phi \lesssim 0.1$ as a conservative bound and leave a more rigorous study of the BBN bounds for future works.

For the case of $\tau_\phi \sim 0.1$~second, the region with $\xi \leqslant \eta_{\mathrm{b}}$ is ruled out since in that region the observed BAU can not be produced before BBN. However, proposals exist for baryogenesis at temperatures as low as $\mathcal{O}(\mathrm{MeV})$ \cite{Ghalsasi:2015mxa,Aitken:2017wie,Elor:2018twp,Nelson:2019fln}; using these models the parameter space with $\xi \leqslant \eta_{\mathrm{b}}$ and $\tau_\phi \lesssim 0.1$~second can still be viable. 

The lower bound on the DM cross sections in Fig.~\ref{fig:WWparameters} for which our analysis is viable comes from the condition that the dark sector has to be thermally populated at early times, assuming a high reheating temperature. 
As we go to lower cross sections, eventually the two sectors will not start from equilibrium and we transition to a freeze-in \cite{Hall:2009bx} scenario instead. Our analytic analysis can be repeated for these cross sections; we leave this study for future work. 

The existence of entropy injection and DM dilution allows DM annihilation cross sections to SM to be below the usual thermal WIMP cross section of Eq.~\eqref{eq:xsecWIMP}. This allows us to have viable models with annihilation cross sections between the conventional freeze-out cross section and the freeze-in cross section. In the heavy DM mass end, this will open up parameter space for DM masses beyond the unitarity bound. At the lower end of the DM mass spectrum, the lower annihilation rates to SM particles lead to open 
parameter space that is accessible to upcoming experiments, 
see Sec.~\ref{sec:experiments} for an example of such models.

\subsection{Generalized freeze-out mass window}
\label{subsec:window}
The entropy injection shifts the required interaction rates for the DM candidates. Using a cross section parametrization, as well as unitarity arguments, this can be translated into a generalized viable DM mass window.

\begin{figure}
    \centering
    \resizebox{0.9\columnwidth}{!}{
    \includegraphics{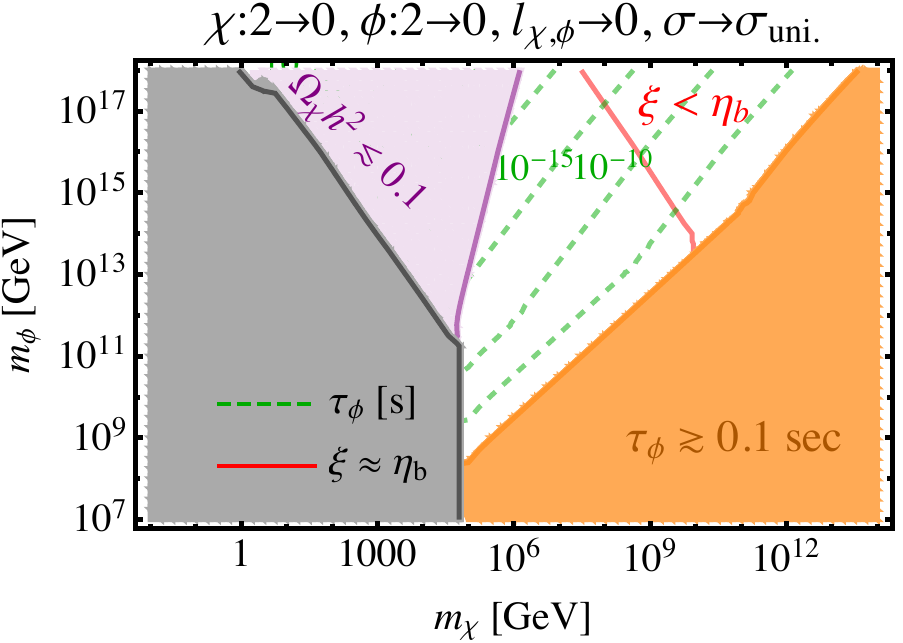} 
    }\\
    \resizebox{0.9\columnwidth}{!}{
    \includegraphics{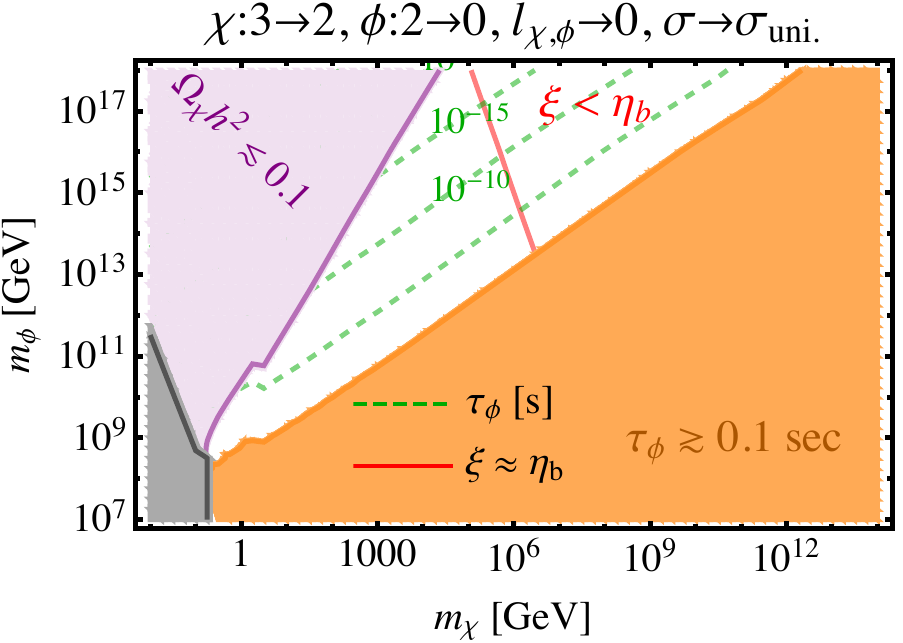} 
    }\\
    \resizebox{0.9\columnwidth}{!}{
    \includegraphics{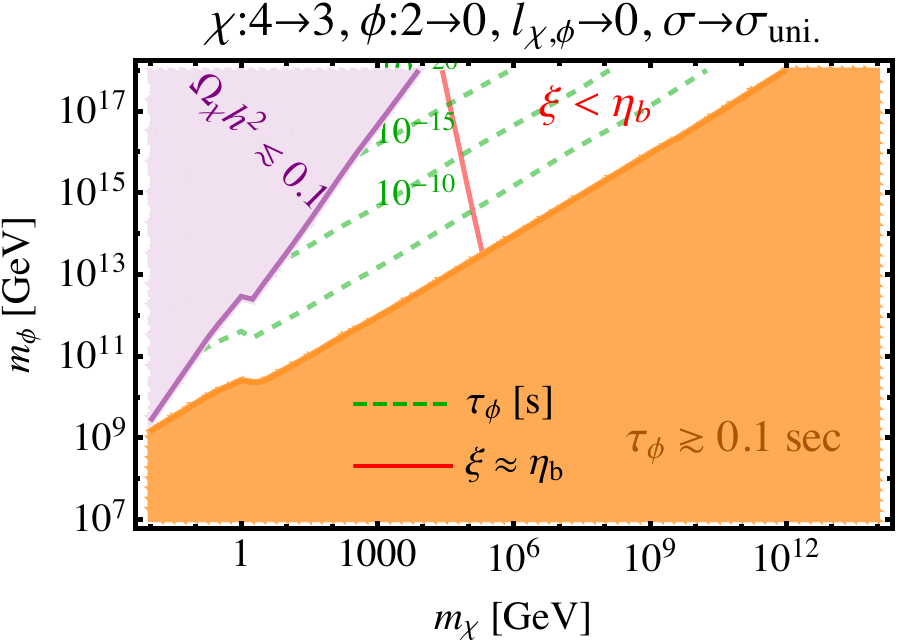} 
    }
    \caption{The viable mass range of $\phi$ and $\chi$ assuming the interactions controlling the $\chi$ freeze-out are $2\rightarrow 0$ (\textbf{top}), $3\rightarrow 2$ (\textbf{middle}), and $4\rightarrow 3$ (\textbf{bottom}), with $\phi$ freezing out via $2\rightarrow 0$ interactions, and no temperature dependence in the freeze-out cross sections. The freeze-out cross sections of both $\chi$ and $\phi$ are set to the maximum value allowed by the unitarity bound for each mass point. We set the lifetime of $\phi$ such that we get the right relic abundance for every point in the parameter space. Contours of a constant lifetime (in seconds) are denoted by dashed green lines. Every point in the orange region requires $\tau_\phi \gsim 0.1$~second to give rise to the observed DM abundance and is ruled out by the BBN bounds. In the gray region, the entropy injection is negligible; the observed DM abundance can be obtained here without any entropy injection and smaller cross sections. On the black line, we get the correct relic abundance with cross sections that saturate the unitarity bound \cite{PhysRevLett.103.153201} and without any entropy injection. In the purple region, if the entropy injection happens after the $\chi$ freeze-out, we will always underproduce DM. For every point to the right of the red line the BAU has to be generated after the $\phi$ entropy injection since in that region $\xi \lesssim \eta_{\mathrm{b}}\sim 6 \times 10^{-10}$. 
    }
    \label{fig:23swaveunitarity}
\end{figure}

In Fig.~\ref{fig:23swaveunitarity}, we show the available parameter space on the plane of $m_\chi-m_\phi$ for different benchmark scenarios. We again use geometric cross sections as estimated upper bounds for the cross sections during freeze-out. Specifically, we take the non-relativistic generalized unitarity bound computed in Eq.~(10) of Ref.~\cite{PhysRevLett.103.153201} and replace the particle momentum with the mass, $k\rightarrow m$:
\begin{equation}
    \langle \sigma v^{p-1} \rangle \lesssim \frac{\Gamma \left(  \frac{3p-3}{2}\right)}{2\pi^{\frac{3p-3}2}} \frac{2\pi p^{\frac{1}{p-1}} p!}{m} \left( \frac{2\pi}{m}\right)^{3p-5} (2l+1).
    \label{eq:unitaritysigma}
\end{equation}
As in the $s$-wave case discussed above, this is justified by the observation that the DM is only mildly non-relativistic during freeze-out; however, if the unitarity bound is truly saturated then the cross sections could be slightly larger than the values given here (and of course much smaller cross sections are always allowed). In App.~\ref{app:morescenarios} we show variations of these plots under rescaling of the assumed maximal cross sections. 

Since the geometric cross section for $\chi$ annihilation should be close to the maximal value allowed by unitarity, using it will allow us to estimate the maximum viable DM mass in each scenario; this is the meaning of the $m_\chi$ axis in Fig.~\ref{fig:23swaveunitarity}. If we go to low enough DM masses, the right relic abundance can be obtained without an entropy injection (and with smaller freeze-out cross sections). This is indicated by the gray region in the plots. In all these plots we set the $\phi$ lifetime for every point in the parameter space such that the final $\Omega_\chi$ matches the observed value \cite{Planck:2018vyg}. 

In each plot, the part with $\tau_\phi$ lifetime longer than $\sim 0.1$~second is ruled out by the BBN bounds \cite{Jedamzik:2006xz,Kawasaki:2017bqm}. 
For large enough hierarchies between $m_\phi$ and $m_\chi$, even if the $\phi$ particles decay right after $\chi$ freezes out, they still inject too much entropy into the SM bath and dilute the DM abundance to below the observed value; thus, in this part of the parameter space, our model can account only for a fraction of DM today.

Clearly, depending on the type of interaction controlling the freeze-outs, the viable mass range shifts significantly. 
In particular, when $p_\chi=2$ (top plot in Fig.~\ref{fig:23swaveunitarity}), without any entropy injection we find the right DM abundance for $m_\chi \sim 100$~TeV (note that the cross sections used in this plot saturate the unitarity bound, thus this agrees with the results of Ref.~\cite{Griest:1989wd}). With entropy injection, we can have viable parameter space with $m_\chi \gsim 100$~TeV. 
For the case of $p_\chi=3$ this bound on the mass in the case of zero entropy injection is shifted to $m_\chi \sim 0.1$~GeV (consistent with Ref.~\cite{Hochberg:2014dra}). For $m_\phi \leqslant 10^{18}$~GeV, in the $p_\chi=2$ ($p_\chi=3$) case the DM mass can go up to even $\sim 10^{14}$~GeV ($\sim 10^{12}$~GeV), provided the unstable relic $\phi$ is long-lived enough.

It is worth reiterating that the final parameter space is not especially sensitive to $q_{\chi,\phi}$ as these quantities merely give rise to $\mathcal{O}(1)$ changes in the final $\Omega_\chi$. 
The plots in Fig.~\ref{fig:23swaveunitarity} are made assuming $l_{\chi,\phi}=0$ interactions; we find that changing these parameters does not affect the viable mass window perceptibly either. 
Results of Sec.~\ref{sec:dilution} can be used to remake these plots with smaller cross sections or different freeze-out interactions for $\chi$ or $\phi$ as well. For completeness, in App.~\ref{app:morescenarios} we include some plots on the viable mass range in a handful of other freeze-out scenarios.

The bottom plot in Fig.~\ref{fig:23swaveunitarity} shows that the entropy injection allows heavy DM masses even for $p_\chi \geqslant 4$ freeze-out interactions. 
Given this enormous viable mass range, it will be interesting to find actual models in which the freeze-out is controlled by $p_\chi \geqslant 4$ interactions. For instance, this can be achieved if we assume the process controlling the freeze-out respects a large discrete symmetry, e.g. $\mathbb{Z}_{N}$ with large enough $N$, between DM particles.

We also assumed there are unspecified 2-to-2 elastic interactions between the SM and each of the relics $\phi$ and $\chi$ that keep them in kinetic equilibrium with the thermal bath. 
The crossing symmetry can relate this elastic process to a number changing process with $p_{\chi,\phi}=2$ that can affect the freeze-out dynamics. 
In Ref.~\cite{Hochberg:2014dra}, however, it was shown that for masses above 1 MeV there can always exist a range of couplings for which kinetic equilibrium can be maintained during the freeze-out via a 2-to-2 elastic scattering while the associated $p_\chi=2$ annihilation process is suppressed.

All in all, Fig.~\ref{fig:23swaveunitarity} shows a vast mass range that can explain the observed DM abundance which, above all parameters, is most sensitive to $p_{\chi,\phi}$. In particular, in such a simple setup the unstable particle $\phi$ can decay even before the electroweak phase transition, i.e. $\tau_\phi \lesssim 10^{-11}$s, and still inject a significant amount of entropy into the SM bath. 
With such a short lifetime, even if the dilution factor is small enough to wash away any pre-existing baryon asymmetry ($\xi \lesssim \eta_\mathrm{b}$), we can regenerate BAU via electroweak baryogenesis proposals \cite{KUZMIN198536}.

Our result shows that even in minimal scenarios including an entropy injection, there is essentially no preferred mass window for thermal relic DM. 
Such a broad mass window became viable thanks to the entropy injection, which in turn takes place if and only if we have an early MD epoch in the universe. Looking for other signatures of such an epoch could be the most model-independent way to probe the high mass end of our viable parameter space, see for example Refs.~\cite{Erickcek:2011us,StenDelos:2019xdk,Barenboim:2021swl}.

\section{viable parameter space in a benchmark model}
\label{sec:experiments}

As a concrete example of how entropy injection expands the parameter space of simple thermal relic DM models, in this section, we study a gauged $\rm B-L$ symmetry with some entropy injection at around $\tau_\phi \approx 0.1$~seconds. 
We do not consider the freeze-out of $\phi$ for simplicity and treat the dilution factor $\xi$ as a free parameter.

We assume the $Z'$ mediator of the $U(1)_{\rm B-L}$ gauge gets a mass via the Stuckelberg mechanism~\cite{Feldman:2011ms}. 
A new vector-like fermion $\chi$ with a $\rm B-L$ charge $n_{\chi} \neq 1$ is then accidentally stable, and provides a minimal DM candidate~\cite{Duerr:2015wfa,Duerr:2015vna}. The dark matter particle $\chi$ freezes out via usual $p_\chi=2$ interactions. 
The relevant interactions are
\begin{align}
     \mathcal{L} \supset - i \gbl n_\chi \bar{\chi} \gamma_\mu \chi Z'^{\mu}  - i \gbl n_f \bar{f} \gamma_\mu f  Z'^{\mu} \, , 
\end{align}
where $n_\chi$ ($n_f$) is the DM (SM fermions) charge under $U(1)_{\rm B-L}$, and $\gbl$ is the gauge coupling of the $U(1)_{\rm B-L}$ gauge group. The charges of the SM particles are $-1$ for leptons and $1/3$ for quarks, such that nucleons couple with a ${\rm B-L}$ charge $n_N =1$. The DM charge will be taken as $n_\chi =3$ for concreteness.

\begin{figure}
    \centering
    \resizebox{0.9\columnwidth}{!}{
    \includegraphics{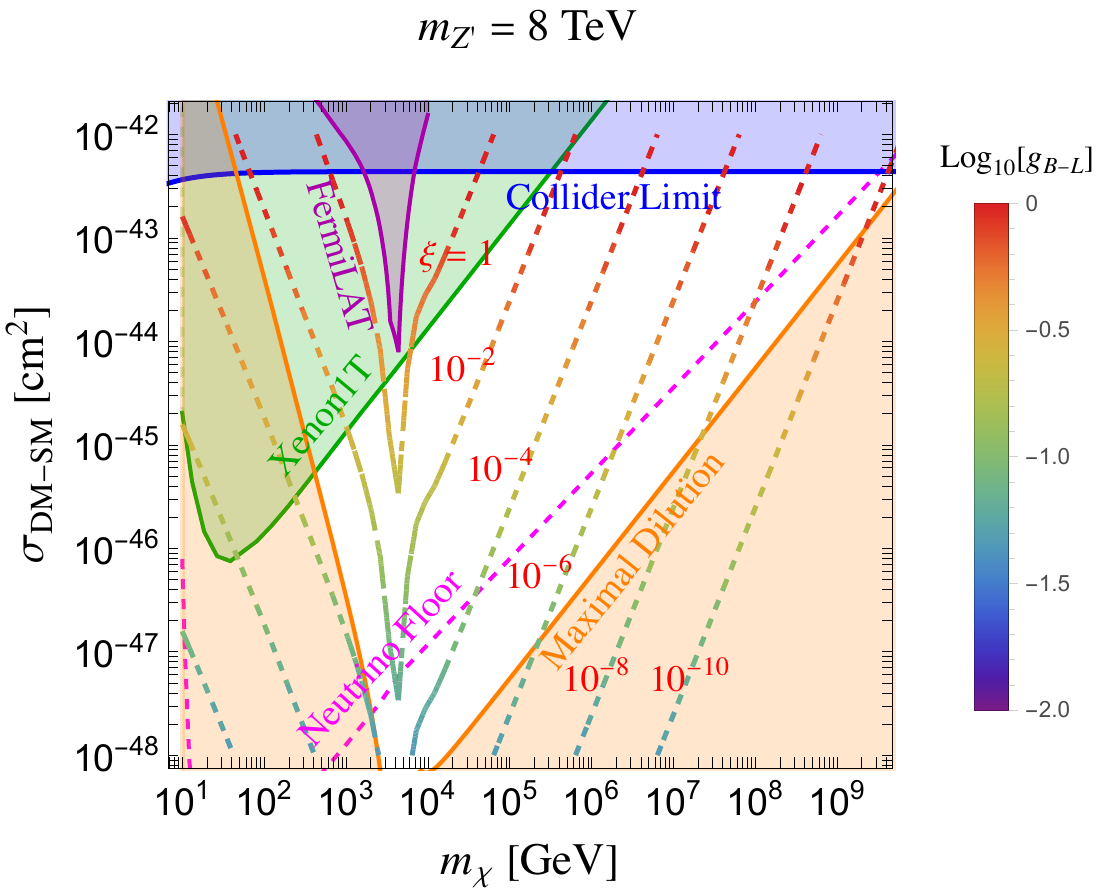}  }\\
  
    \caption{Contours of required dilution factor $\xi$ to generate the observed DM abundance today on the plane of DM mass vs. spin-independent DM-nucleon cross section in our benchmark $U(1)_{\mathrm{B-L}}$ mediator model. We fix the mediator mass to $m_{Z'}=8$~TeV and focus on the heavy DM region.  Contours of constant $\xi$ are color-coded according to the value of $\gbl$. 
    Superimposed are the bounds from collider searches for the $\mathrm{B-L}$ gauge boson~\cite{BaBar:2014zli} (blue), current direct detection bounds from Xenon1T~\cite{XENON:2018voc} (green), the indirect DM annihilation limits from the FermiLAT satellite~\cite{Fermi-LAT:2017opo} (purple), the expected neutrino floor~\cite{Billard:2013qya} (dashed magenta), and maximal dilution limit if $\chi$ freezes out during a RD epoch (orange), see the text for further details.  }
    \label{fig:HeavyDM}
\end{figure}

\begin{figure}
    \centering
    \resizebox{0.9\columnwidth}{!}{
    \includegraphics{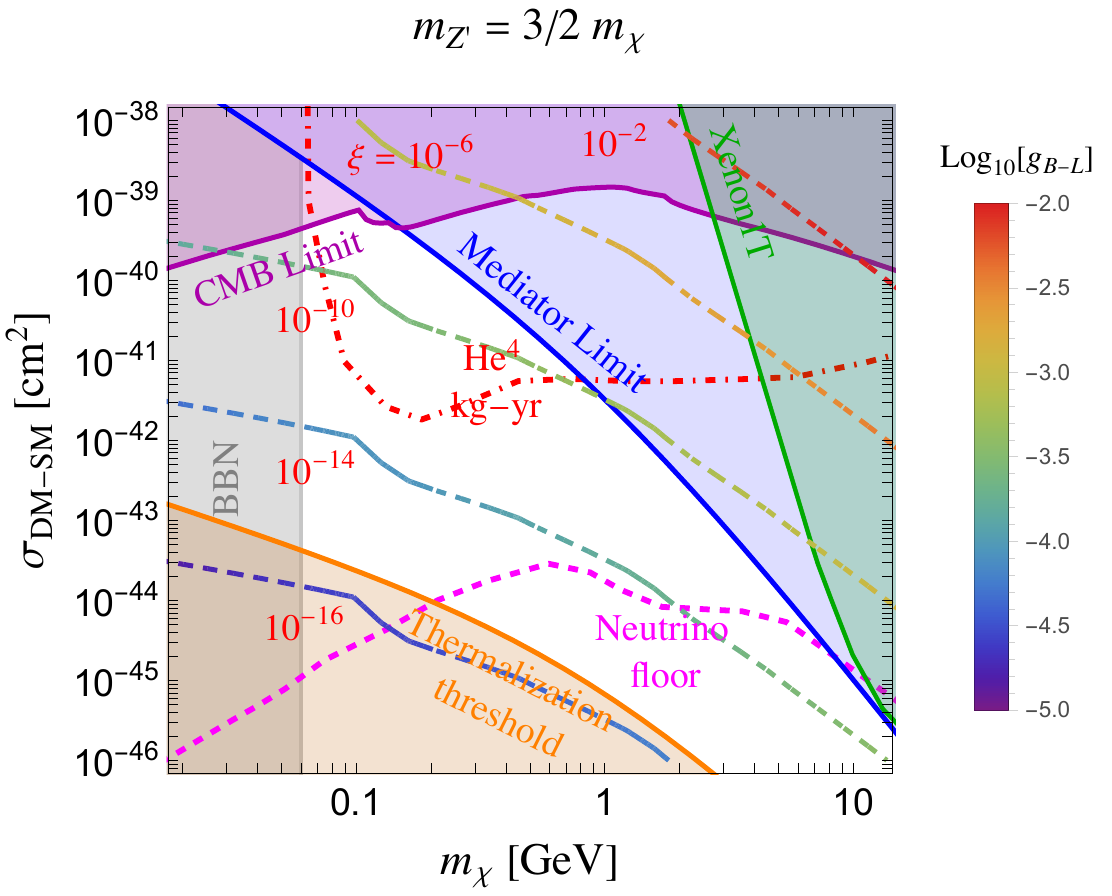} }
    \caption{The parameter space in the light DM regime. Similar color-coding as in Fig.~\ref{fig:HeavyDM} is used for the constant $\xi$ contours. 
    We show the current Xenon1T~\cite{XENON:2018voc} bounds (green), expected reach of low threshold superfluid helium searches~\cite{Hertel:2018aal} (red dot-dashed contour), the collider bounds on the $\mathrm{B-L}$ mediator from BaBar~\cite{BaBar:2014zli} (see also \cite{Heeck:2014zfa}) (blue), indirect DM annihilation limits from Planck observations~\cite{Slatyer:2015jla} (purple), BBN bounds (gray), thermalization threshold (orange), and expected neutrino floor~\cite{Billard:2013qya} (dashed magenta). The entropy injection opens new viable parameter space for such light DM candidates; see the text for further details.  
    }
    \label{fig:lightDM}
\end{figure}

In Fig.~\ref{fig:HeavyDM} and Fig.~\ref{fig:lightDM}, we show the plane of dark matter mass vs. spin-independent elastic cross section. We study both limits of heavy (above the $\sim 100 \, \rm TeV$ unitarity bound) and light (sub-GeV) DM masses. 
The $U(1)_{\mathrm{B-L}}$ gauge coupling is uniquely determined for any point on this plot. We then choose the right dilution factor $\xi$ for each point such that the right DM abundance today is obtained; contours of required $\xi$ are shown. 

In Fig.~\ref{fig:HeavyDM}, we show the heavy DM scenario with a fixed mediator mass. The indirect detection limit from the FermiLAT satellite~\cite{Fermi-LAT:2017opo} is only relevant for the case that no dilution is present, i.e. $\xi =1$. However, as discussed in Refs. \cite{Erickcek:2011us,StenDelos:2019xdk,Barenboim:2021swl}, the early MD epoch can lead to increased production of high-density DM mini-halos. This effect could lead to an enhanced DM annihilation signal, if those halos are not disrupted at later times.  In particular in the case of a low reheating temperature, this effect could improve the indirect detection prospects by several orders of magnitude.

The LEP collider bound of $m_{Z'}/\gbl \lesssim 6.9 \, \rm TeV$~\cite{ALEPH:2013dgf,Heeck:2014zfa} provides an upper bound on the spin-independent cross section $\sigma_{\rm DM-SM} \lesssim 5 \cdot 10^{-44} \, n_\chi^2 \, \text{cm}^2$. The current Xenon1T limit \cite{XENON:2018voc} excludes the no-dilution scenario in a wide range of masses as well.

For simplicity, in making this plot we assumed $\chi$ freezes out during a RD epoch, i.e. $T_i \leqslant T_\chi$. Since the entropy injection should occur before $t\sim 0.1$~second, i.e. $\Gamma_\phi \geqslant \left( 0.1 ~ \mathrm{second} \right)^{-1}$, we can calculate a lower bound on $\xi$ using Eq.~\eqref{eq:EntDump}, see the orange region in Fig.~\ref{fig:HeavyDM}. Combining this limit with the collider searches mentioned above puts an upper bound of $m_\chi \lesssim 10^{10}$~GeV on DM mass in this benchmark model.

With this example we demonstrate that for $\xi \ll 1$ a wide parameter space remains open for DM masses well above the perturbative unitarity bound, and significant gauge interactions $\gbl \sim \mathcal{O}(1)$. Such scenarios without the weakness of the WIMP couplings are within the reach of upcoming large volume detectors, such as XENONnT~\cite{XENON:2020kmp}, and provide excellent experimental targets. We also find viable parameter space below the neutrino floor~\cite{Monroe:2007xp,Vergados:2008jp,Billard:2013qya} which motivates developing new search techniques, see for example Refs.~\cite{Hochberg:2016ntt,Essig:2016crl,Kadribasic:2017obi,Budnik:2017sbu,Rajendran:2017ynw,Griffin:2018bjn,Coskuner:2019odd,Blanco:2021hlm}. 

In Fig.~\ref{fig:lightDM}, we show contours of required $\xi$ to get the right DM abundance today in the light DM regime.  We assume the DM freeze-out happens during a MD epoch and we fix $\tau_\phi  = 0.1$~second, such that the entropy injection happens right before the BBN epoch. Note that in most of the viable parameter space $\xi < \eta_{\mathrm{b}}$; in those parameter regions the BAU should be generated either after the entropy injection, e.g. using models from Refs.~\cite{Ghalsasi:2015mxa,Aitken:2017wie,Elor:2018twp,Nelson:2019fln}, or by $\phi$ decay itself.

Superimposed on this figure are the limits from searches for the $\mathrm{B-L}$ gauge boson in the sub-GeV mass range at BaBar \cite{BaBar:2014zli} (see also Ref.~\cite{Heeck:2014zfa}), as well as the limit on the $s$-wave annihilation cross section from the CMB~\cite{Slatyer:2015jla}, and current Xenon1T bounds~\cite{XENON:2018voc}. 
At very small coupling values the system never enters thermal equilibrium with the SM and our treatment is not applicable. The lower $m_\chi$ end of the plot is ruled out by BBN constraints.

Overall, while the no-dilution scenario is ruled out in this DM mass range, a large parameter space with $\xi \lesssim 10^{-8}$ is widely open, motivating new search strategies that can cover the entire viable region. 
We find that part of the remaining parameter space is within the reach of upcoming low threshold direct detection techniques with superfluid helium~\cite{Knapen:2017xzo,Hertel:2018aal} with one kg-year exposure; larger low threshold detectors can fully test the low mass end of this thermal DM scenario within the next decade~\cite{Knapen:2017xzo}.

\section{Conclusion and Outlook}
\label{sec:conclusion}

We studied the freeze-out abundance of thermal relic DM (denoted $\chi$) in a next-to-minimal freeze-out scenario featuring an unstable relic $\phi$ that decays to the SM after its freeze-out. We analytically calculated the final $\chi$ abundance for general freeze-out interactions with $p_{\chi,\phi} \geqslant 2$ initial particles and $q_{\chi,\phi} \geqslant 0$ final particles and for arbitrary temperature dependence of the freeze-out cross sections. 

We found that the final $\chi$ abundance is very sensitive to $m_{\chi,\phi}$, their cross sections, and especially the initial number of particles in the interaction ($p_{\chi,\phi}$). The viable DM mass window moves as $p_{\chi,\phi}$ changes, even in the presence of entropy injection. However, even for the conventional WIMP scenario ($(p_\chi,~q_\chi)=(2,0)$) the entropy injection means there is no particular mass value singled out -- unlike the traditional WIMP models in which the weak scale emerges as the natural DM mass scale. 

Using our analytic calculation we identified viable parameter space between the freeze-in and freeze-out cross section values (Fig.~\ref{fig:WWparameters}). We also pointed out that for the WIMP scenario, if the BAU is generated after (before) the entropy injection, the DM can be as heavy as $\sim 10^{14}~$GeV ($\sim 10^{8}~$GeV) for $m_\phi \lesssim M_{\mathrm{Pl}}$ as shown in Fig.~\ref{fig:23swaveunitarity}. 
We also used our results to argue that for freeze-out interactions with $p_\chi \geqslant 3$ there is viable parameter space with DM masses as high as $m_\chi \sim 10^{12}$~GeV.

As a proof of principle, we studied the parameter space of a simple $\mathrm{B-L}$ gauge boson portal and showed that the entropy injection can open up parameter space for light, and heavy DM masses that were previously considered ruled out owing to direct detection bounds for the former and CMB bounds for the latter ( Figs.~\ref{fig:HeavyDM}-\ref{fig:lightDM}). 

Our calculation and analysis can be extended in a few interesting ways.
It is intriguing to consider the case of either the $\chi$ or $\phi$ relic density being set by a freeze-in instead of freeze-out. In particular, this way we can extend the range of Fig.~\ref{fig:WWparameters} to lower cross sections in each direction. Furthermore, given the large viable masses that will not be accessible to conventional DM searches in foreseeable future, the best way to probe our setup is likely by looking for signatures of an early MD epoch. There are already proposals in the literature for searching for an early MD epoch \cite{Erickcek:2011us,StenDelos:2019xdk,Barenboim:2021swl}. Such an epoch can also affect the evolution of the Hubble parameter and thus the spectrum of gravitational waves from various sources in the early universe \cite{Cui:2018rwi,Auclair:2019wcv,Gouttenoire:2019kij,Hook:2020phx,Chang:2021afa}.

There are also intriguing studies of UV-complete scenarios which we leave for future explorations. In particular, now that we have a large parameter space for freeze-out via $p_\chi \rightarrow q_\chi$ interactions with $p_\chi \geqslant 4$, it is interesting to think about natural particle physics models that could give rise to such freeze-out behavior. 

Another interesting direction will be studying the correlation between DM and SM abundances in more detail. One of the best arguments for the existence of a portal between the DM and the SM is the closeness of their respective abundances. Since the SM abundance today is set by the BAU, a complete study should include a way to relate this asymmetry to the DM abundance today. Since the DM abundance in our setup is a function of the $\phi$ lifetime, a natural direction for exploration is UV-complete models where the BAU is generated by decays of $\phi$ itself.

\section*{Acknowledgments}

We thank Zackaria Chacko, John Beacom, Rouven Essig, Aditya Parikh, Matt Reece, Filippo Sala, Kai Schmidt-Hoberg, and Raman Sundrum for useful discussions. The work of PA and TRS was supported by the U.S. Department of Energy, Office of Science, Office of High Energy Physics, under grant Contract Number DE-SC0012567. \\

\appendix


\section{Entropy injection}
\label{app:entropydump}

The relevant calculation for the entropy injection is carried out for a single heavy relic decaying to the SM in Ref.~\cite{Kolb:1990vq} section 5.3. (Similar calculation is carried out elsewhere in the literature as well; see for instance Refs.~\cite{Berlin:2016gtr,Contino:2018crt}.) For completeness we repeat that calculation here; we also generalize that to the case where the dark sector has a different temperature than SM. While this is not the case in our study, this can be useful for future works.

We want to calculate the amount of entropy injected into SM when $\phi$ decays into SM bath. First, we should keep in mind that the energy density of $\phi$ is controlled by
\begin{equation}
\frac{1}{R^3}\frac{d}{dt} \left( \rho_\phi R^3 \right) = \dot{\rho}_\phi + 3 H \rho_\phi = - \Gamma_\phi \rho_\phi,
\label{eq:GBdecayeq}
\end{equation}
with $\rho_\phi$ being $\phi$'s energy density, $\Gamma_\phi$ their decay rate, and $H$ the Hubble constant. The solution to this equation is 
\begin{equation}
\rho_\phi (R) = \rho (R_0) \frac{R_0^3}{R^3} e^{-\Gamma_\phi t},
\label{eq:GBdecaysol}
\end{equation}
where the subscript $0$ here refers to some arbitrary origin of time (and not today) and $t$ is the time passed since then. This equation and solution is valid for both RD and MD portion of the evolution, as long as $\phi$ is non-relativistic.

As the $\phi$ particles decay to SM, they inject heat and thus entropy into it. The entropy injection can be calculated by:
\begin{equation}
dS = \frac{dQ}{T} = - \frac{4\pi}{3} \frac{d(R^3 \rho_\phi)}{T} = \frac{4\pi}{3} \frac{R^3}{T} \rho_\phi \Gamma_\phi dt,
\label{eq:dS}
\end{equation}
where in the second equality we used Eq.~\eqref{eq:GBdecayeq}. Notice that $dS$ here is the infinitesimal change in the entropy of SM. The entropy of the SM can be written as
\begin{equation}
S = \frac{4\pi R^3}{3} \frac{2\pi^2}{45} g_{*S} T^3,
\label{eq:Sdef}
\end{equation}
where $T$ and $g_{*S}$ refer to SM temperature and number of relativistic degrees of freedom ($\#$dofs) for entropy. Without any entropy injection from decoupled particles, this total entropy of SM remains constant. Putting the last two equations together, we have
\begin{equation}
\dot{S} S^{1/3} = \left(\frac{4\pi}{3}\right)^{4/3} R^4 \rho_\phi (R) \Gamma_\phi \left(	\frac{2\pi^2}{45} \right)^{1/3} g_{*S}^{1/3}.
\label{eq:Scoreeq}
\end{equation}

\begin{widetext}
This equation can be solved as
\begin{eqnarray}
\int S^{1/3} dS &=& \left(\frac{4\pi}{3}\right)^{4/3} \left(	\frac{2\pi^2}{45} \right)^{1/3}  \int dt R^4 \rho_\phi (R) \Gamma_\phi g_{*S}^{1/3} \nonumber \\
\label{eq:Ssolvformal}
\Longrightarrow S_f^{4/3} &=& S_i^{4/3} + \frac{4}{3} \left(\frac{4\pi}{3}\right)^{4/3} \left(	\frac{2\pi^2}{45} \right)^{1/3}\int_{t_i}^{t_f} dt R^4 \rho_\phi (R) \Gamma_\phi g_{*S}^{1/3} ,
\end{eqnarray}
where now $i,f$ are just two labels for the beginning and end of any interval we are studying. Combining this with Eq.~\eqref{eq:GBdecaysol}, we find
\begin{equation}
S_f^{4/3} = S_i^{4/3} + \frac{4}{3} \left(\frac{4\pi}{3}\right)^{4/3} \left(	\frac{2\pi^2}{45} \right)^{1/3} \rho_i R_i^4 \Gamma_\phi \int_{t_i}^{t_f} dt \frac{R}{R_i} g_{*S}^{1/3} e^{-\Gamma_\phi t}
\label{eq:Sfsolsimp1}
\end{equation}
\begin{equation}
\Longrightarrow \frac{S_f}{S_i} = \left(1 +  \frac{4}{3} \left(\frac{4\pi}{3}\right)^{4/3}  \frac{\rho_i}{S_i^{4/3}} R_i^4 \mathcal{I} \right)^{3/4},
\label{eq:Sfsolsimp2}
\end{equation}
where
\begin{equation}
\mathcal{I} = \Gamma_\phi \left(	\frac{2\pi^2}{45} \right)^{1/3} \int_{t_i}^{t_f} dt \frac{R}{R_i} g_{*S}^{1/3} e^{-\Gamma_\phi t}.
\label{eq:def-I}
\end{equation}
We can simplify this quantity as below
\begin{eqnarray}
\mathcal{I} &=& \Gamma_\phi \frac{\int_{t_i}^{t_f} dt  \left(	\frac{2\pi^2}{45} \right)^{1/3} \frac{R}{R_i} g_{*S}^{1/3} e^{-\Gamma_\phi t}}{\int_{t_i}^{t_f} dt \left(	\frac{2\pi^2}{45} \right)^{1/3} \frac{R}{R_i}  e^{-\Gamma_\phi t}} \int_{t_i}^{t_f} dt \left(	\frac{2\pi^2}{45} \right)^{1/3} \frac{R}{R_i}  e^{-\Gamma_\phi t}\nonumber \\
&\equiv & \Gamma_\phi <g_{*S}^{1/3}>  \int_{t_i}^{t_f} dt \left(	\frac{2\pi^2}{45} \right)^{1/3} \frac{R}{R_i}  e^{-\Gamma_\phi t}.
\label{eq:gS*avg}
\end{eqnarray}
\end{widetext}
From this point on, again for simplicity we assume $i$ labels a point right after the universe becomes MD, i.e. $\rho_{\phi,i} \approx \rho_{R,i}$ with $\rho_{R}$ denoting the radiation bath energy density. When the universe becomes MD, and before $\phi$ decays ($t \ll \tau_\phi $), we have 
\begin{eqnarray}
\frac{R}{R_i} &=& \left(	\frac{\rho_{\phi,i}}{\rho_\phi}	\right)^{1/3}, \\ \nonumber
\label{eq:RRiratio}
\rho_\phi &=& \frac{3 M_{pl}^2}{8\pi} \left(	\frac{\dot{R}}{R}	\right)^2, \\ 
\Longrightarrow \frac{R}{R_i} & =& \rho_{\phi,i}^{1/3} \left(	\frac{8\pi}{3 M_{pl}^2}	\right)^{1/3} \left(	\frac{\dot{R}}{R}	\right)^{-2/3}. \nonumber
\end{eqnarray}
In the MD universe, we have $R \propto t^{2/3(1+w)}$ with $w=0$, i.e. $R \propto t^{2/3}$. Hence, in the MD epoch, we have
\begin{equation}
\left(\frac{\dot{R}}{R} \right)^{-2/3} = \left( \frac{3}{2} t \right)^{2/3}.
\label{eq:Rdotscale}
\end{equation}
\begin{widetext}
Combining these equations, we can rewrite Eq.~\eqref{eq:gS*avg} as
\begin{eqnarray}
\label{eq:Isimplified}
\mathcal{I} &=& <g_{*S}^{1/3}>  \Gamma_\phi  \int_{t_i}^{t_f} dt \left(	\frac{2\pi^2}{45} \right)^{1/3} \rho_{\phi,i}^{1/3} \left(	\frac{8\pi}{3 M_{pl}^2}	\right)^{1/3} \left( \frac{3}{2} t \right)^{2/3} e^{-\Gamma_\phi t} \\
&=&<g_{*S}^{1/3}> \left(	\frac{2\pi^2}{45} \right)^{1/3} \rho_{\phi,i}^{1/3} \left( \frac{3}{2} \right)^{2/3} \left(	\frac{8\pi}{3 M_{pl}^2}	\right)^{1/3} \Gamma_\phi^{-2/3}  \int_{u_i}^{u_f} du   u^{2/3} e^{-u},~~u=\Gamma_\phi t. \nonumber
\end{eqnarray}

For $t_i \ll \tau_\phi$, we can simply use $t_i=0$. However, we should be careful about the upper bound of the integral. As mentioned above, we are considering a MD universe. 
So, the equations are really only reliable as long as most of the $\phi$s have not decayed. Yet, let us for the moment assume the universe actually does remain MD even after the $\phi$ decays. 
This introduces some error since the $R/R_i$ scaling changes after $\phi$ decays, but with this assumption, we can take $t_f \rightarrow \infty$ and use the definitions of the Gamma functions:
\begin{eqnarray}
\label{eq:Ifinals}
\mathcal{I} &  = &  <g_{*S}^{1/3}> \left(	\frac{2\pi^2}{45} \right)^{1/3} \rho_{\phi,i}^{1/3} \left( \frac{3}{2} \right)^{2/3} \left(	\frac{8\pi}{3 M_{pl}^2}	\right)^{1/3} \Gamma_\phi^{-2/3}  \Gamma(5/3) \\
&\approx & 1.18    \rho_{\phi,i}^{1/3} \left(	\frac{8\pi}{3 M_{pl}^2}	\right)^{1/3} \left(	\frac{2\pi^2}{45} \right)^{1/3} <g_{*S}^{1/3}> \Gamma_\phi^{-2/3} ,
\end{eqnarray}
which is off by only around $10\%$ from Eq.~(5.72) of \cite{Kolb:1990vq}, where the integral is evaluated numerically with the right scaling of $R$ in the RD epoch after our MD era. We can now use this in Eq.~\eqref{eq:Sfsolsimp2} to find
\begin{equation}
\frac{S_f}{S_i} = \left(	1 + \frac{4}{3} \left(	\frac{4\pi}{3}	\right)^{4/3}	\frac{\rho_{\phi,i}}{S_i^{4/3}} R_i^4 \times  1.18    \rho_{\phi,i}^{1/3} \left(	\frac{8\pi}{3 M_{pl}^2}	\right)^{1/3} \left(	\frac{2\pi^2}{45} \right)^{1/3} <g_{*S}^{1/3}> \Gamma_\phi^{-2/3} \right)^{3/4} .
\label{eq:result1}
\end{equation}

Note should be taken that $g_{*}$ and $S_{i,f}$ are referring to SM quantities. The entropy and the energy density of SM at $t_i$ can be written as
\begin{eqnarray}
\label{eq:SMstuffs}
S_i &=& \frac{2\pi^2}{45} g_{*S,i} T_i^3 \frac{4\pi R_i^3}{3}, \\
\rho_i &=& \frac{\pi^2}{30} g_{*,i} T_i^4, \nonumber \\
& \Longrightarrow  &\frac{4}{3} \left(	\frac{4\pi}{3}	\right)^{4/3}	\frac{\rho_{\phi,i}}{S_i^{4/3}} R_i^4 = \left(  \frac{45}{2\pi^2} \right)^{1/3} \frac{g_{*,i}}{g_{*S,i}^{4/3}} \\
 \Longrightarrow   S_f/S_i & = & \left( 1 +  \frac{g_{*,i}}{g_{*S,i}^{4/3}}  <g_{*S}^{1/3}> 1.18 \rho_{\phi,i}^{1/3} \left(	\frac{8\pi}{3 M_{pl}^2}	\right)^{1/3} \Gamma_\phi^{-2/3}  \right)^{3/4} \\
& = & \left( 1 + 2.40 \frac{g_{*,i}}{g_{*S,i}^{4/3}}  <g_{*S}^{1/3}> \frac{}{} \rho_{\phi,i}^{1/3} \left(	\frac{1}{M_{pl}^2}	\right)^{1/3} \Gamma_\phi^{-2/3}  \right)^{3/4}.
\end{eqnarray}
Finally, again using $\rho_{\phi,i} \approx \rho_{R,i} = \pi^2/30 \times g_{*,i} T_i^4$, we find
\begin{equation}
\frac{S_f}{S_i} = \left( 1+ 1.65 \left(\frac{g_{*,i}}{g_{*S,i}}\right)^{4/3} <g_{*S}^{1/3}> \left(\frac{T_i^4}{(\Gamma_\phi M_{pl} )^{2}} \right)^{1/3} \right)^{3/4}.
\label{eq:finalSSratio}
\end{equation}
\end{widetext}

The only place where the temperature difference of the two sectors could enter the calculation is in $\#$dofs $g_{*(S)}$. Notice that all these $\#$dofs factors entered from energy or entropy density of SM and did not care about $\#$dofs in the dark sector. 
Thus, \textit{even if the two sectors had different temperatures, the calculation would have remained unchanged.} Furthermore, since all particles are at the same $T$, $g_{*}=g_{*S}$, 
\begin{equation}
\frac{S_f}{S_i} = \left( 1+ 1.65 <g_{*S}^{1/3}> \left(\frac{T_i^4}{(\Gamma_\phi M_{pl} )^{2}} \right)^{1/3} \right)^{3/4}.
\label{eq:finalSfactor}
\end{equation}
which is the final formula we use in Sec.~\ref{sec:parameters}. The $g_{*}$ average is defined in Eq.~\eqref{eq:gS*avg} and given its small power, it will always be an $\mathcal{O}(1)$ number.

\section{Details of freeze-out and Dilution During a RD Epoch}
\label{app:RD}

Let us assume the process controlling the freeze-out is $p \rightarrow q + \mathrm{SM}$, i.e. $p$ initial DM particles go to $q$ final DM particles and an arbitrary number of particles from the SM thermal bath. The calculation here is viable for both $\phi$ and $\chi$ relics, so we drop their subscripts in this appendix. 
Assuming identical particles and anti-particles, the Boltzmann equation for this freeze-out is:
\begin{eqnarray}
\dot{n}&+& 3Hn = s\dot{Y} = - \frac{(p-q)}{p!} \langle \sigma v^{p-1}\rangle (n^p - n_{\mathrm{eq}}^{p-q} n^q) \nonumber \\
\label{eq:pqfoequation}
&=& - \frac{(p-q)}{p!} \langle \sigma v^{p-1}\rangle s^p Y^p \left(1 - \left(\frac{Y_{\mathrm{eq}}}{Y}\right)^{p-q}\right).
\end{eqnarray}
In a RD universe, we have (see Eq.~(5.15) of \cite{Kolb:1990vq})
\begin{equation}
t = 0.301 g_*^{-1/2}\frac{M_{\mathrm{Pl}}}{T^2} \equiv \frac{x^2}{2H(m)},
\label{eq:tinRD}
\end{equation}
where $m$ is the relic that freezes out, $x= m/T$, and $H(m)=1.67 g_*^{1/2} m^2/M_{\mathrm{Pl}}$. This suggests that in RD epochs:
\begin{equation}
\frac{\partial x}{ \partial t} = \frac{H(m)}{x}.
\label{eq:RDdxdt}
\end{equation}
Thus, the Boltzmann equation can be rewritten as
\begin{equation}
Y' = - \frac{p-q}{p!} \frac{x}{H(m)} \langle \sigma v^{p-1} \rangle s^{p-1} Y^p \left(	1 - \left(\frac{Y_{\mathrm{eq}}}{Y}\right)^{p-q}	\right).
\label{eq:boltzRD2}
\end{equation}
We also have
\begin{equation}
s = \frac{2\pi^2}{45} g_{*S} T^3 \equiv \beta x^{-3},
\label{eq:RDsdef}
\end{equation}
using which we can rewrite the Boltzmann equation as
\begin{equation}
Y' = - \frac{p-q}{p!} \frac{x^{-3p+4}}{H(m)} \langle \sigma v^{p-1} \rangle \beta^{p-1} Y^p \left(	1 - \left(\frac{Y_{\mathrm{eq}}}{Y}\right)^{p-q}	\right).
\label{eq:boltzRD3}
\end{equation}
We can use this equation to find an approximation for $x_{\mathrm{f.o.}}$, i.e. when the freeze-out happens (see below).

When calculating the asymptotic yield, $Y_{\mathrm{eq}} \ll Y$, we have:
\begin{equation}
\frac{dY}{dx} = - \frac{p-q}{p!} \frac{x^{-3p+4}}{H(m)} \langle \sigma v^{p-1} \rangle \beta^{p-1} Y^p.
\label{eq:boltzRD4}
\end{equation}
At this point we have to determine the $x$ dependence of the cross section. We simply assume
\begin{equation}
\langle \sigma v^{p-1} \rangle \equiv  \sigma_0 \sum_i a_i x^{-i}.
\label{eq:xsectionexpansion}
\end{equation}
This captures a general temperature dependence for the interaction. 
Then Eq.~\eqref{eq:boltzRD4} can be rewritten as
\begin{equation}
\frac{dY}{Y^p} = - dx \frac{p-q}{p!} \frac{\beta^{p-1}}{H(m)} \left(   \sigma_0 \sum_i a_i x^{-i-3p+4} \right) .
\label{eq:boltzRD5}
\end{equation}
Integrating this equation we find the asymptotic value of the abundance $Y_\infty$: 
\begin{widetext}
\begin{eqnarray}
\label{eq:solallpartial}
&\frac{1}{(p-1)Y^{p-1}_\infty} &= \frac{p-q}{p!} \frac{\beta^{p-1}}{H(m)} \left( \sigma_0 \sum_i \frac{a_i}{(3p+i-5)x_{\mathrm{f.o.}}^{3p+i-5} }\right) \\
\label{eq:solallpartial2}
&\Longrightarrow & Y_\infty = \left(\frac{p!}{(p-q)(p-1)}\right)^{\frac{1}{p-1}}\frac{H(m)^{\frac{1}{p-1}}}{\beta} \left( \frac{1}{\sigma_0\sum_i \frac{a_i}{(3p+i-5)x_{\mathrm{f.o.}}^{3p+i-5}}  } \right)^{\frac{1}{p-1}} .
\end{eqnarray}
Notice that in this integration we neglected the change in $g_{*(S)}$ during the freeze-out. The only place the RD property of the universe enters is in the $H(m)$ factor and the the derivative $\partial t /\partial x$. Replacing $H(m)$ and $\beta =  \frac{2\pi^2}{45} g_{*S} m^3$ in the expression above we find:
\begin{eqnarray}
\label{eq:solallpartial3}
Y_\infty &=& \frac{45 }{2\pi^2} \left(\frac{p!}{(p-q)(p-1)}\right)^{\frac{1}{p-1}}  \left(   1.67 g_*^{1/2}  \right)^{\frac{1}{p-1}} \times \frac{m^{\frac{2}{p-1}-3}}{g_{*S}} \left( \frac{1}{\sigma_0 M_{\mathrm{Pl}} \sum_i \frac{a_i}{(3p+i-5)x_{\mathrm{f.o.}}^{3p+i-5}}  } \right)^{\frac{1}{p-1}} ,
\end{eqnarray}
where $g_{*(S)}$ is the $\#$dofs (for entropy) during the freeze-out, i.e. what we called $g_{*S,\chi}$ or $g_{*S,\phi}$ in the main body of the paper. 
If we assumed $\langle \sigma v\rangle = \sigma_0 x^{-l}$ instead, the expression above simplifies to:
\begin{eqnarray}
\label{eq:solonepartial}
Y_\infty &=& \frac{45 }{2\pi^2} \left(\frac{p!}{(p-q)(p-1)}\right)^{\frac{1}{p-1}}  \left(   1.67 g_*^{1/2}  \right)^{\frac{1}{p-1}} 
\times  \frac{m^{\frac{2}{p-1}-3}}{g_{*S}} \left( \frac{(3p+l-5)x_{\mathrm{f.o.}}^{3p+l-5} }{\sigma_0 M_{\mathrm{Pl}} } \right)^{\frac{1}{p-1}} , 
\end{eqnarray}
which allows us to calculate $\Omega h^2$:
\begin{eqnarray}
\label{eq:finalOmega}
\Omega h^2 &=& m Y_\infty \frac{s_0}{\rho_c}h^2 \\
&=& \frac{s_0}{\rho_c}h^2 \times \frac{45 }{2\pi^2} \left(\frac{p!}{(p-q)(p-1)}\right)^{\frac{1}{p-1}}  \left(   1.67 g_*^{1/2}  \right)^{\frac{1}{p-1}} \frac{m^{\frac{2}{p-1}-2}}{g_{*S}} \left( \frac{(3p+l-5)x_{\mathrm{f.o.}}^{3p+l-5} }{\sigma_0 M_{\mathrm{Pl}} } \right)^{\frac{1}{p-1}}. \nonumber 
\end{eqnarray}

Following Ref.~\cite{Kolb:1990vq}, we define the freeze-out point $x_{\mathrm{f.o.}}$ as when 
$\partial_x (Y-Y_{\mathrm{eq}})=0$ and $Y=(1+c)Y_{\mathrm{eq}}$ with $c$ being an $\mathcal{O}(1)$ number. Assuming a non-relativistic distribution for the $Y_{\mathrm{eq}}$, we can algebraically solve these equations for $x_{\mathrm{f.o.}}$ to find
\begin{eqnarray}
    \label{eq:xfo}
    x_{\mathrm{f.o.}} &\approx &\frac{1}{p-1} \ln \left( \left( 0.14 \times\frac{ \mathrm{g}}{g_{*S}} \right)^{p-1} \lambda (1+c)^p \right) \\ 
    &-& \frac{3p+2l-5}{2(p-1)} \ln \left[ \frac{1}{p-1} \ln \left( \left( 0.14 \times\frac{ \mathrm{g}}{g_{*S}} \right)^{p-1} \lambda (1+c)^{p} \right) \right], \nonumber
\end{eqnarray}
where
\begin{eqnarray}
\label{eq:xfoparameters}
\lambda  & = &  \frac{p-q}{p !} \frac{\sigma_{0}}{H(m)} \beta^{p-1}, 
\end{eqnarray}
\end{widetext}
$\mathrm{g}$ is the internal degrees of freedom of the particle that freezes out, and we used $\langle \sigma v\rangle = \sigma_0 x^{-l}$.
Varying the value of $c$ only changes $x_{\mathrm{f.o.}}$ by a few percent for $p_\chi \leqslant 10$, which in turn only affects the final abundance by $\mathcal{O}(1)$ factors. We will use $c=1$ in this work. In deriving this equation we again neglected the variation in $\#$dofs during the freeze-out.

\section{Details of freeze-out and Dilution During a MD Epoch}
\label{app:MD}

Now let us consider the scenario where the $\phi$ relic freezes out during a RD epoch, while the $\chi$ relic freezes out after the universe enters the MD epoch. Notice that for the entropy injection to affect the $\chi$ abundance, $\tau_\phi$ should be long enough so that it decays after the $\chi$ freeze-out, i.e. $T_i \gsim T_\chi \gsim T_{\tau_\phi}$. 
The calculation of $\phi$ freeze-out is unchanged from the previous appendix, since it occurs in a RD universe, see Eq.~\eqref{eq:solallpartial3}. 
The energy density of $\phi$ is 
\begin{eqnarray}
\label{eq:MDrhophiappx}
\rho_\phi (T) &=& m_\phi Y_\infty^\phi s(T) \equiv \gamma^2 T^3, \\
\gamma^2 &=& \frac{g_{*S}(T)}{g_{*S\phi}} \left(\frac{p_\phi !}{(p_\phi-q_\phi)(p_\phi-1)}\right)^{\frac{1}{p_\phi-1}}  \left(   1.67 g_{*\phi}^{1/2}  \right)^{\frac{1}{p_\phi-1}} \nonumber \\ 
&\times & m_\phi^{\frac{2}{p_\phi-1}-2} \left( \frac{1}{\sigma_{0,p_\phi} M_{\mathrm{Pl}} \sum_i \frac{a_i}{(3p_\phi+i-5)x_{\mathrm{f.o.},\phi}^{3p_\phi+i-5}}  } \right)^{\frac{1}{p_\phi-1}} \nonumber \\
& \approx  &\left(\frac{p_\phi  !}{(p_\phi-q_\phi)(p_\phi-1)}\right)^{\frac{1}{p_\phi-1}}  \left(   1.67 g_{*\phi}^{1/2}  \right)^{\frac{1}{p_\phi-1}} \nonumber \\
&\times & m_\phi^{\frac{2}{p_\phi-1}-2} \left( \frac{1}{\sigma_{0,p_\phi} M_{\mathrm{Pl}} \sum_i \frac{a_i}{(3p_\phi+i-5)x_{\mathrm{f.o.},\phi}^{3p_\phi+i-5}}  } \right)^{\frac{1}{p_\phi-1}} . \nonumber
\end{eqnarray}
In the last line we neglect the change in $g_{*S}$ after the $\phi$ freeze-out (and effectively until the $\chi$ freeze-out). This was done to make $\xi$ temperature-independent so as to simplify the upcoming integrations. One can show that neglecting this change in $\#$dofs, we have
\begin{equation}
    \gamma^2 \approx \frac{\pi^2}{30} g_{*S,\phi} T_i .
    \label{eq:gammadefappx}
\end{equation}

Now let us study $\chi$'s freeze-out. The temperature evolution in this MD epoch is given by:
\begin{eqnarray}
\label{eq:HinMD}
\partial_t T=-HT&,&~ H= \sqrt{\frac{8\pi \rho_\phi}{3M_{\mathrm{Pl}}^2}},~\rho_\phi = \g^2 T^3 \\
&\Longrightarrow & H = \sqrt{\frac{8\pi}{3}} \frac{\g}{M_{\mathrm{Pl}}} T^{3/2} \nonumber \\
&\Longrightarrow& \frac{dT}{T^{5/2}} = - \sqrt{\frac{8\pi}{3}} \frac{\g}{M_{\mathrm{Pl}}} dt \nonumber \\
&\Longrightarrow & -\frac{2}{3}T^{-3/2} \Big|_i^f = - \sqrt{\frac{8\pi}{3}} \frac{\g}{M_{\mathrm{Pl}}} t \Big|_i^f , \nonumber
\end{eqnarray}
and we can neglect the initial $T$ and $t$ since time and temperature change significantly during the integration; thus
\begin{eqnarray}
\Longrightarrow  t &=& \frac{2}{3}T^{-3/2} \sqrt{\frac{3}{8\pi}} \frac{M_{\mathrm{Pl}}}{\g} \equiv \frac{2}{3} \frac{x_\chi^{3/2}}{H_{\mathrm{MD}}(m_\chi)}, \nonumber \\
\label{eq:HinMD2}
H_{\mathrm{MD}}(m_\chi)&=& \sqrt{\frac{8\pi}{3}} \g  \frac{m_\chi^{3/2}}{M_{\mathrm{Pl}}}. 
\end{eqnarray}
This is the equivalent of Eq.~\eqref{eq:tinRD} for a MD universe; we see that $t$ has a different scaling with $T$ or $x_\chi$ now. The following equations for $\chi$ freeze-out in the previous section can now be rewritten as below 
\begin{equation}
\label{eq:dxdtMD}
\frac{\partial x}{\partial t} = \frac{H_{\mathrm{MD}}(m_\chi)}{x_\chi^{1/2}}.
\end{equation}
Thus, the Boltzmann equation can be rewritten as
\begin{widetext}
\begin{eqnarray}
\label{eq:boltzMD2}
Y' &=& - \frac{p_\chi-q_\chi}{p_\chi !} \frac{x_\chi^{1/2}}{H_{\mathrm{MD}}(m_\chi)} \langle \sigma v^{p_\chi-1} \rangle s^{p_\chi-1} Y^{p_\chi} \left(	1 - \left(\frac{Y_{\mathrm{eq}}}{Y}\right)^{p_\chi-q_\chi}	\right)\\
\label{eq:boltzMD3}
&=& - \frac{p_\chi-q_\chi}{p_\chi !} \frac{x_\chi^{-3p_\chi+\frac{7}{2}}}{H_{\mathrm{MD}}(m_\chi)} \langle \sigma v^{p_\chi-1} \rangle \beta^{p_\chi-1} Y^{p_\chi} \left(	1 - \left(\frac{Y_{\mathrm{eq}}}{Y}\right)^{p_\chi-q_\chi}	\right)\\
\label{eq:boltzMD4}
& \approx & - \frac{p_\chi-q_\chi}{p_\chi !} \frac{x_\chi^{-3p_\chi+\frac{7}{2}}}{H_{\mathrm{MD}}(m_\chi)} \langle \sigma v^{p_\chi-1} \rangle \beta^{p_\chi-1} Y^{p_\chi},
\end{eqnarray}
where again in the last line we have considered the asymptotic form of the equation when $Y_\mathrm{eq} \ll Y$. Compared to the RD case, the power of $x_\chi$ and the $H_\mathrm{MD}(m_\chi)$ parameter have changed.

Using the general formula for the cross section Eq.~\eqref{eq:xsectionexpansion}, we now can have

\begin{equation}
\frac{dY}{Y^{p_\chi}} = - dx_\chi \frac{p_\chi-q_\chi}{p_\chi !} \frac{\beta^{p_\chi-1}}{H_{\mathrm{MD}}(m_\chi)} \left(   \sigma_{0,p_\chi} \sum_i a_i x^{-i-3p_\chi+7/2} \right) .
\label{eq:boltzMD5}
\end{equation}
Once we integrate this equation from $x=x_{\mathrm{f.o.}}$ to $x \rightarrow \infty$,  
we find
\begin{eqnarray}
\label{eq:solallpartialMD}
\frac{1}{(p_\chi-1)Y^{p_\chi-1}_\infty} &=& \frac{p_\chi-q_\chi}{p_\chi !} \frac{\beta^{p_\chi-1}}{H_{\mathrm{MD}}(m_\chi)} \left( \sigma_{0,p_\chi} \sum_i \frac{a_i}{(3p_\chi+i-9/2)x^{3p_\chi+i-9/2} }\right) \\
\label{eq:solallpartial2MD}
\Longrightarrow Y_\infty^\chi &=& \left(\frac{p_\chi !}{(p_\chi-q_\chi)(p_\chi-1)}\right)^{\frac{1}{p_\chi-1}}\frac{H_{\mathrm{MD}}(m_\chi)^{\frac{1}{p_\chi-1}}}{\beta} \left( \frac{1}{\sigma_{0,p_\chi}\sum_i \frac{a_i}{(3p_\chi+i-9/2)x^{3p_\chi+i-9/2}}  } \right)^{\frac{1}{p_\chi-1}} .
\end{eqnarray}

Again we neglected the variation in $\#$dofs in the integration above. (This can introduce some error for DM mass in the $\mathcal{O}(0.1)$-$\mathcal{O}(10)$~GeV ballpark; for this mass range, the calculation should be carried out numerically for better precision.) 
Now we should replace $H_{\mathrm{MD}}(m_\chi)$ from Eq.~\eqref{eq:HinMD2} and $\beta$ from Eq.~\eqref{eq:RDsdef} (for $\chi$). 
The final abundance of $\chi$ becomes
\begin{eqnarray}
\label{eq:solallpartial3MD}
Y_\infty^\chi &=& \frac{45 }{2\pi^2} \left(\frac{p_\chi !}{(p_\chi-q_\chi)(p_\chi-1)}\right)^{\frac{1}{p_\chi-1}}  \left(   2.89 \gamma  \right)^{\frac{1}{p_\chi-1}}  \frac{m^{\frac{3/2}{p_\chi-1}-3}}{g_{*S,\chi}} \left( \frac{1}{\sigma_0 M_{\mathrm{Pl}} \sum_i \frac{a_i}{(3p_\chi+i-9/2)x^{3p_\chi+i-9/2}}  } \right)^{\frac{1}{p_\chi-1}} . 
\end{eqnarray}
This is the equivalent of Eq.~\eqref{eq:solallpartial3} in a MD epoch freeze-out. Specializing to the simpler form of the cross section $\langle \sigma v^{p_\chi-1} \rangle \equiv \sigma_{0,p_\chi} x_\chi^{-l_\chi}$, we find
\begin{eqnarray}
\label{eq:solonepartialMD}
Y_\infty^\chi &=& \frac{45 }{2\pi^2} \left(\frac{p_\chi !}{(p_\chi-q_\chi)(p_\chi-1)}\right)^{\frac{1}{p_\chi-1}}  \left(   2.89 \gamma   \right)^{\frac{1}{p_\chi-1}} \nonumber \frac{m^{\frac{3/2}{p_\chi-1}-3}}{g_{*S,\chi}} \left( \frac{(3p_\chi+l_\chi-9/2)x^{3p_\chi+l_\chi-9/2} }{\sigma_{0,p_\chi} M_{\mathrm{Pl}} } \right)^{\frac{1}{p_\chi-1}} . 
\end{eqnarray}
This equation is in agreement with the results of Ref.~\cite{Hamdan:2017psw}. 
To turn this yield into $\Omega_\chi h^2$ for a massive relic, we use:
\begin{eqnarray}
\label{eq:finalOmegaMDappx}
\Omega_\chi h^2 &=& m_\chi Y_\infty^\chi \frac{s_0}{\rho_c} h^2  \\
&=& \frac{s_0}{\rho_c}h^2 \times \frac{45 }{2\pi^2} \left(\frac{p_\chi !}{(p_\chi-q_\chi)(p_\chi-1)}\right)^{\frac{1}{p_\chi-1}}  \left(    2.89 \gamma   \right)^{\frac{1}{p_\chi-1}}  \frac{m_\chi^{\frac{3/2}{p_\chi-1}-2}}{g_{*S,\chi}} \left( \frac{(3p_\chi+l_\chi-9/2)x_\chi^{3p_\chi+l_\chi-9/2} }{\sigma_{0,p_\chi} M_{\mathrm{Pl}} } \right)^{\frac{1}{p_\chi-1}}. \nonumber
\end{eqnarray}
\end{widetext}
By replacing the value of $\gamma$ we find the final expression used in Sec.~\ref{subsec:FOinMD}.

We can also use Eq.~\eqref{eq:boltzMD3} to find an expression for $x_{\mathrm{f.o.}}$ during a MD epoch as well. In this case, we find 
\begin{eqnarray}
\label{eq:xfoMD}
x_{\mathrm{f.o.}} &=& \mathrm{Eq}.~\eqref{eq:xfo} \Big|_{l_\chi \rightarrow l_\chi + 1/2}^{H (m_\chi) \rightarrow H_{\mathrm{MD}} (m_\chi)} .
\end{eqnarray}

\section{Additional Scenarios}
\label{app:morescenarios}

\begin{figure}
    \centering
    \resizebox{0.9\columnwidth}{!}{
    \includegraphics{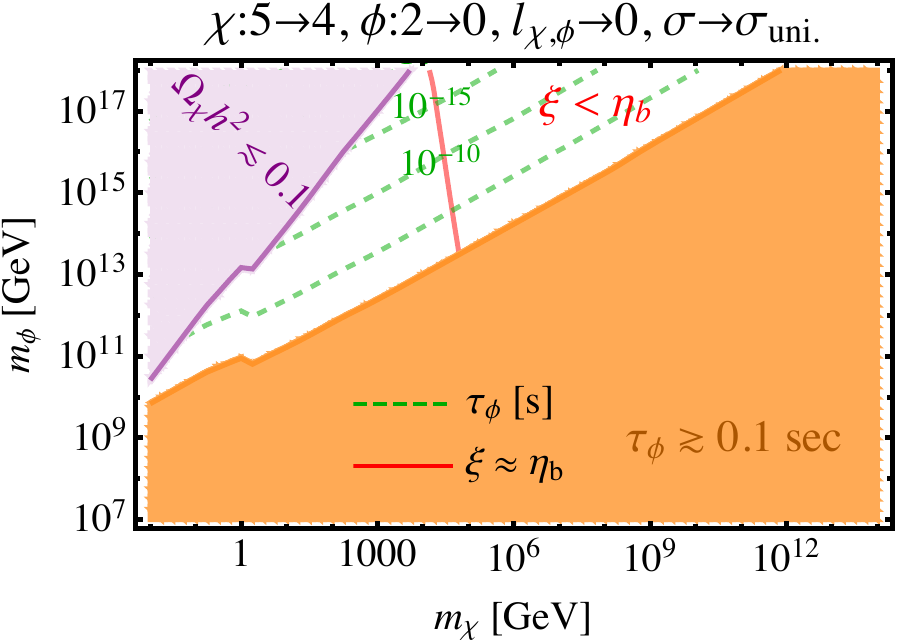} 
    }
    \caption{ Similar to Fig.~\ref{fig:23swaveunitarity} but with $p_\chi=q_\chi+1=5$. We still find a large viable parameter space.
    }
    \label{fig:5scenariosfigs}
\end{figure}

\begin{figure*}
    \centering
    \resizebox{1.8\columnwidth}{!}{
    \includegraphics{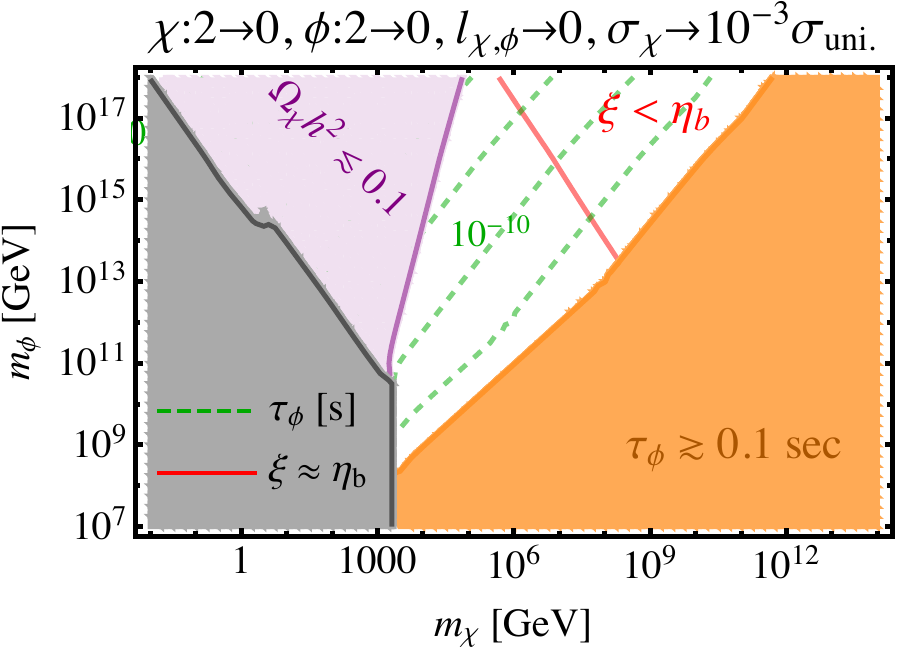} 
    \includegraphics{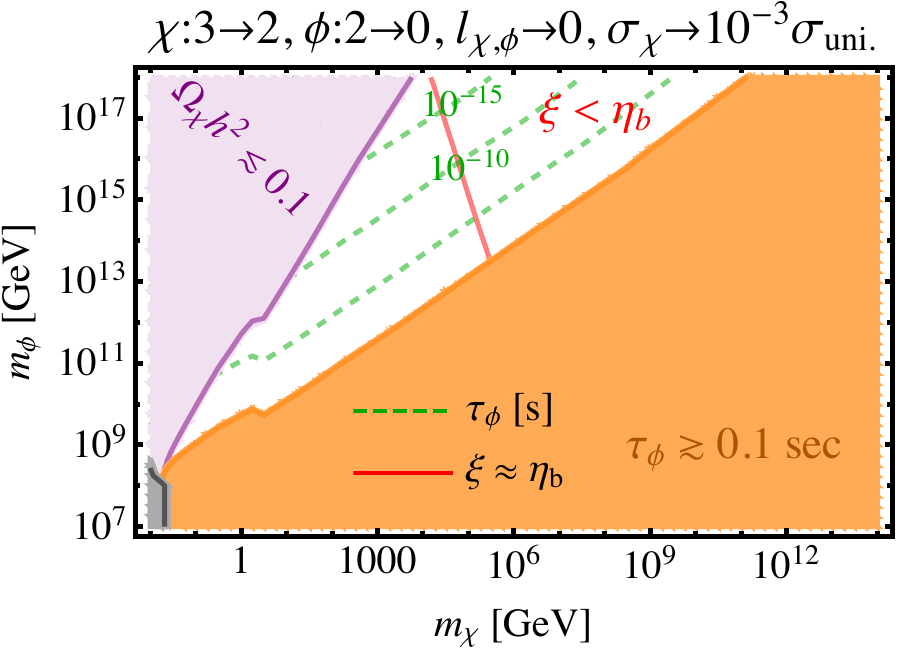} 
    }
    \caption{ Similar to Fig.~\ref{fig:23swaveunitarity} but with smaller $\sigma_{0,\chi}$ values. Lowering this freeze-out cross section slightly shifts the viable parameter space to lower $m_\chi$ values.
    }
    \label{fig:nonunitaryscenariosfigs}
\end{figure*}

\begin{figure*}
    \centering
    \resizebox{1.8\columnwidth}{!}{
    \includegraphics{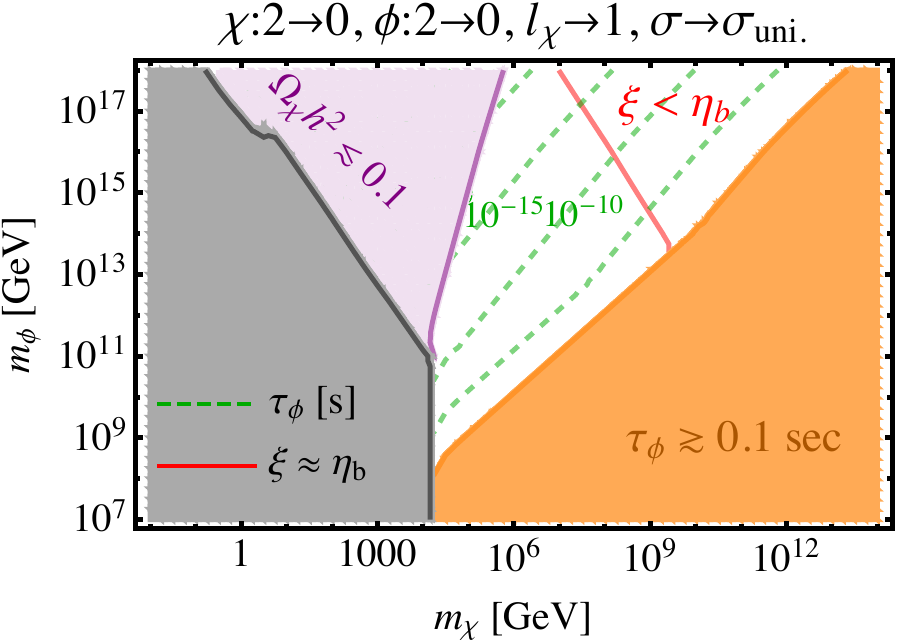} 
    \includegraphics{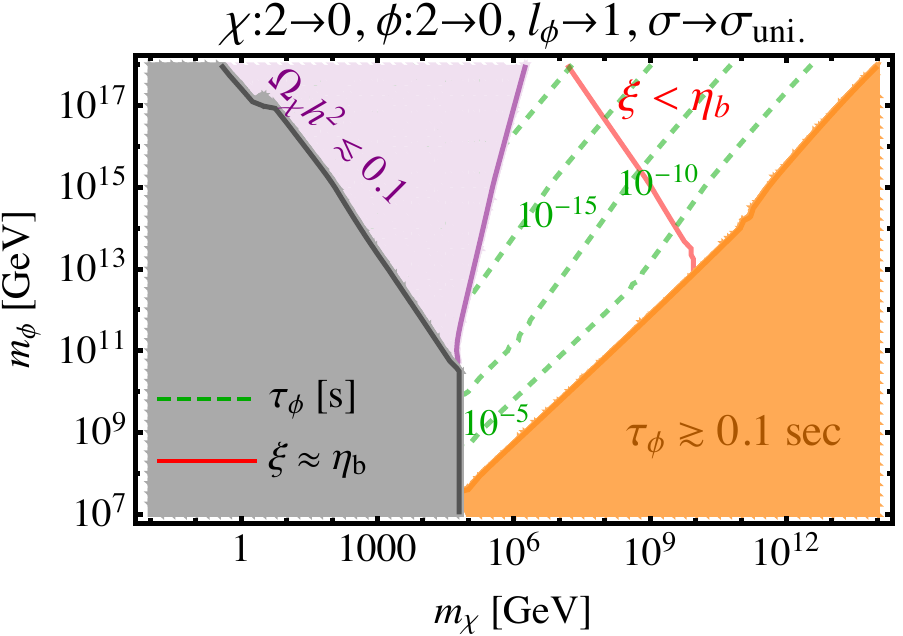} 
    }
    \caption{ Similar to Fig.~\ref{fig:23swaveunitarity} but with different $l_{\chi}$ and $l_{\phi}$ values. We find that increasing each of these quantities slightly shifts the parameter space to lower values of $m_\chi$ or $m_\phi$, respectively. We note that the effect of changing $l_{\chi,\phi}$ is much smaller than changing $p_{\chi,\phi}$, via comparison to the top plot of Fig.~\ref{fig:23swaveunitarity}.
    }
    \label{fig:spfigs}
\end{figure*}

\begin{figure*}
    \centering
    \resizebox{1.8\columnwidth}{!}{
    \includegraphics{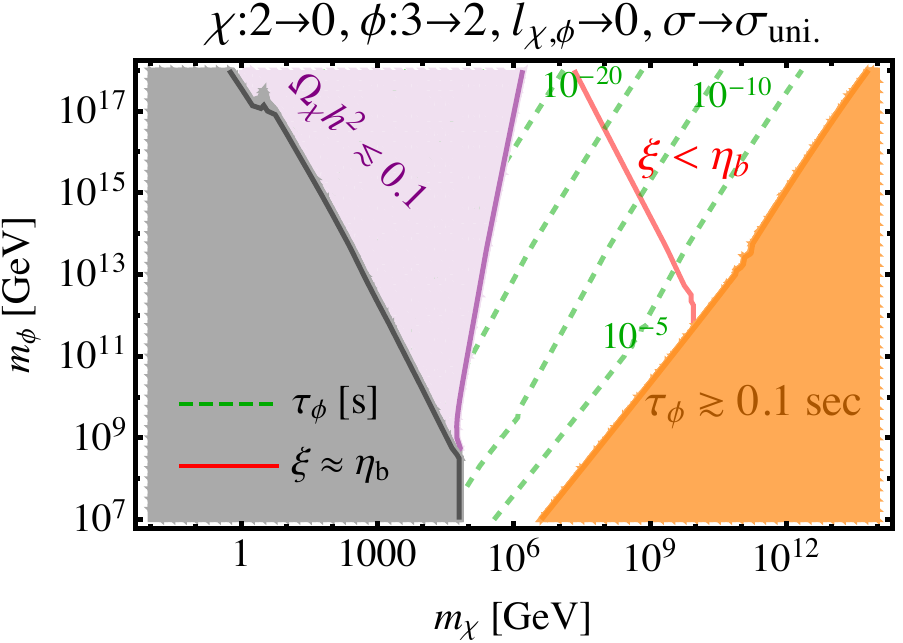} 
    \includegraphics{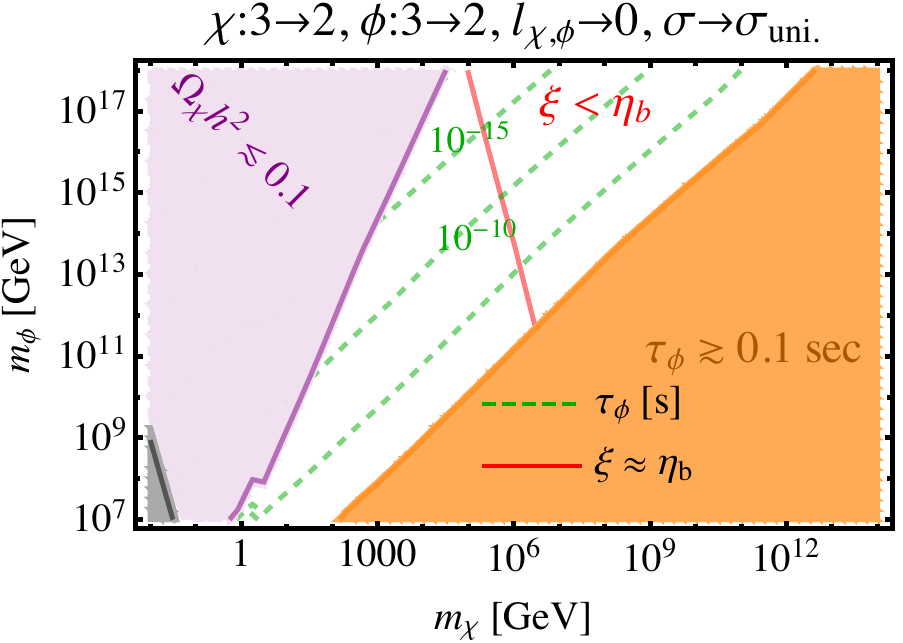} 
    }
    \caption{ Similar to Fig.~\ref{fig:23swaveunitarity} but with $p_\phi=q_\phi+1=3$ and different $p_{\chi}$ and $q_{\chi}$ values. Increasing $p_\phi$ significantly shifts the open parameter space to lower $m_\phi$ values (compared to Fig.~\ref{fig:23swaveunitarity}).
    }
    \label{fig:3scenariosfigs}
\end{figure*}

In the body of the paper we reported the viable mass window and $\Gamma_\phi$ values for a few $(p_{\chi, \phi},q_{\chi, \phi},l_{\chi, \phi})$ scenarios and assuming geometric cross sections. Our general formulas in Eqs.~\eqref{eq:finalpqdilutedresult} and \eqref{eq:finalpqdilutedresultMD} can be used for other values of these quantities as well. We include a few more mass window plots in Figs.~\ref{fig:5scenariosfigs}-\ref{fig:3scenariosfigs}; comparing these figures to Fig.~\ref{fig:23swaveunitarity} can provide us with intuition on the effect of various parameters on the viable DM mass range.

We show the viable mass window for the case of $p_\chi=q_\chi+1=5$ in Fig.~\ref{fig:5scenariosfigs}; we find a large viable parameter space similar to the case of $p_\chi=q_\chi+1=4$, see the bottom plot of Fig.~\ref{fig:23swaveunitarity}.

In the top (bottom) panel of Fig.~\ref{fig:nonunitaryscenariosfigs} we show the viable mass range and $\Gamma_\phi$ values with $(p_{\chi,\phi},q_{\chi,\phi},l_{\chi,\phi})=(2,0,0)$ ($(p_{\chi,\phi},q_{\chi,\phi},l_{\chi,\phi})=(3,2,0)$), similar to the top (middle) panel of Fig.~\ref{fig:23swaveunitarity}) but with a cross section much smaller than the geometric cross section for $\chi$ particles. 
As expected, decreasing the cross section shifts the viable parameter space to lower $m_\chi$ values.

In Fig.~\ref{fig:spfigs} we keep $(p_{\chi,\phi},q_{\chi,\phi})=(2,0)$ and use the geometric cross sections for each freeze-out but vary the value of $l_{\chi, \phi}$. 
We find that increasing $l_\chi$ ($l_\phi$) slightly shifts the viable parameter space to lower $m_\chi$ ($m_\phi$) values. 
However, the effect of $l_{\chi,\phi}$ on the viable mass range is clearly not as strong as that of $p_{\chi,\phi}$.

These plots can be compared to those of Fig.~\ref{fig:3scenariosfigs} where we keep $l_{\chi,\phi}=0$ and instead change $p_{\chi,\phi}$. 
We again observe a shift to lower $m_\chi$ or $m_\phi$ values, respectively, when $p_\chi$ or $p_\phi$ increases; the effect of changing $p_{\chi,\phi}$ is clearly stronger than changing $l_{\chi,\phi}$ in Fig.~\ref{fig:spfigs}.

\clearpage

\bibliographystyle{utphys_modified}
\bibliography{bib}

\end{document}